\documentclass[12pt,preprint]{aastex}
\usepackage{graphicx}
\input epsf

\newcommand{\euve}{{\it EUVE}}
\newcommand{\deltE}{\Delta\kern-1ptE}
\newcommand{\fuse}{{\it FUSE}}
\newcommand{\hst}{{\it HST}}
\newcommand{\kms}{km~s$^{-1}$}
\newcommand{\gal}{$\alpha$}
\newcommand{\gb}{$\beta$}
\newcommand{\gla}{$\lambda$}
\newcommand{\etal}{et al.}
\newcommand{\ecs}{erg cm$^{-2}$s$^{-1}$}

\begin{document}

\title{A Far Ultraviolet Spectroscopic Survey of Luminous Cool Stars}
\author{A. K. Dupree, A. Lobel, and  P. R. Young\altaffilmark{1}}
\affil{Harvard-Smithsonian Center for Astrophysics, \\
Cambridge MA 02138}
\email{adupree@cfa.harvard.edu, alobel@cfa.harvard.edu, p.r.young@rl.ac.uk}

\and

\author{T. B. Ake}
\affil{Johns Hopkins University/CSC\\
Baltimore, MD 21218}
\email{ake@pha.jhu.edu}

\and

\author{J. L. Linsky, and S. Redfield\altaffilmark{2}}
\affil{JILA, University of Colorado and NIST \\
Boulder, CO 80309-0440}
\email{jlinsky@jila.colorado.edu, sredfiel@astro.as.utexas.edu}

\begin{abstract}
\fuse\ ultraviolet spectra of 8 giant and supergiant 
stars reveal that high temperature (3$\times$10$^5$K)
atmospheres are common in luminous cool stars and extend 
across the color-magnitude diagram 
from \gal~Car (F0 II) to the cool giant \gal\ Tau (K5 III).  
Emission present 
in these spectra includes chromospheric H-Ly\gb, Fe~II, C~I, and 
transition region lines of  
C~III, O~VI, Si III, Si~IV. Emission lines of Fe XVIII and
Fe XIX signaling temperatures of $\sim$10$^7$K and coronal material
are found in
the most active stars, \gb\ Cet and 31 Com.  A short-term flux
variation, perhaps a flare, was detected in \gb\ Cet during our
observation.  Stellar surface fluxes
of the emission of C III and O VI are correlated and  decrease rapidly 
towards the cooler stars, reminiscent of the decay of magnetically-heated atmospheres. 
Profiles of the C III (\gla 977) lines  suggest that   
mass outflow is underway
at T$\sim$80,000 K, and the winds are warm. 
Indications of outflow at higher temperatures (3$\times$10$^5$K) are
revealed by O VI asymmetries and the line widths themselves.  High temperature
species are absent in the M-supergiant \gal~Ori.  
Narrow fluorescent lines of Fe II appear in the spectra
of many giants and supergiants,
apparently pumped by H Ly-\gal, and formed in extended atmospheres.
Instrumental characteristics that affect cool star spectra are discussed.
\end{abstract}
\altaffiltext{1}{Currently
at Rutherford Appleton Laboratory, Culham UK}
\altaffiltext{2}{Currently at McDonald Observatory, University
of Texas, Austin, TX 78712-1083}

\section{Introduction}
The structure  of the outer atmospheres of cool giant and supergiant
stars can reveal the evolution of magnetic activity as atmospheres
expand, stars lose angular momentum, and arguably 
dynamo heating decreases, 
as stars become cooler and more luminous.  The existence 
of hot material and its relation to winds and mass loss
can be addressed with 
far ultraviolet spectra obtained
with the Far Ultraviolet Spectroscopic Explorer (\fuse) 
satellite (Moos \etal\ 2000) because they
probe 
both the presence of high temperature plasma and the
dynamics of the atmosphere.  Two well-recognized examples of
these extremes are represented by the Sun 
(possessing a hot, fast, low mass flux
wind) and the supergiant \gal\ Ori (possessing a cool, slow,
high mass flux wind).  However identifying the connecting
links between these two types of atmospheres, perhaps represented
in part by the hybrid stars (Hartmann \etal\  
1980, Reimers \etal\ 1996),
can be addressed by \fuse\ spectra.  Thus, far ultraviolet
spectra can be used to build a comprehensive
picture of the heating and dynamics of the outer atmospheres of cool stars.

Ultraviolet measurements with the International Ultraviolet Explorer ({\it IUE}) 
laid the foundations for characterization of cool
star atmospheres (cf. Jordan \& Linsky 1989; Dupree \& Reimers
1989); the Hubble Space Telescope  ({\it HST}) has focused principally on individual 
objects (cf Ayres \etal\ 1998, McMurry \& Jordan 2000, Carpenter \etal\ 1999,
Robinson \etal\ 1998, and Lobel \& Dupree 2001).
\fuse\  complements {\it IUE} and {\it HST} because coverage
of shorter wavelengths (\gla\gla912--1180) gives 
access to the
strong \ion{O}{6} resonance emission formed 
at temperatures $\sim 3\times 10^5$K
providing a  diagnostic of temperatures higher than normally
available in the near ultraviolet; and line
profiles provide clues to heating and atmospheric dynamics.  This
spectral region also contains fine structure transitions of \ion{Fe}{18}
and \ion{Fe}{19} that enable detection of a hot corona and its
dynamics and extend
the temperature coverage by more than an order of magnitude
to 7$\times$10$^6$K in addition to allowing velocity and
profile measurements.  A summary of the major atomic transitions
specifically considered here is contained in Table~\ref{tbl.lines}

\begin{figure}
\includegraphics[angle=0,scale=1.0]{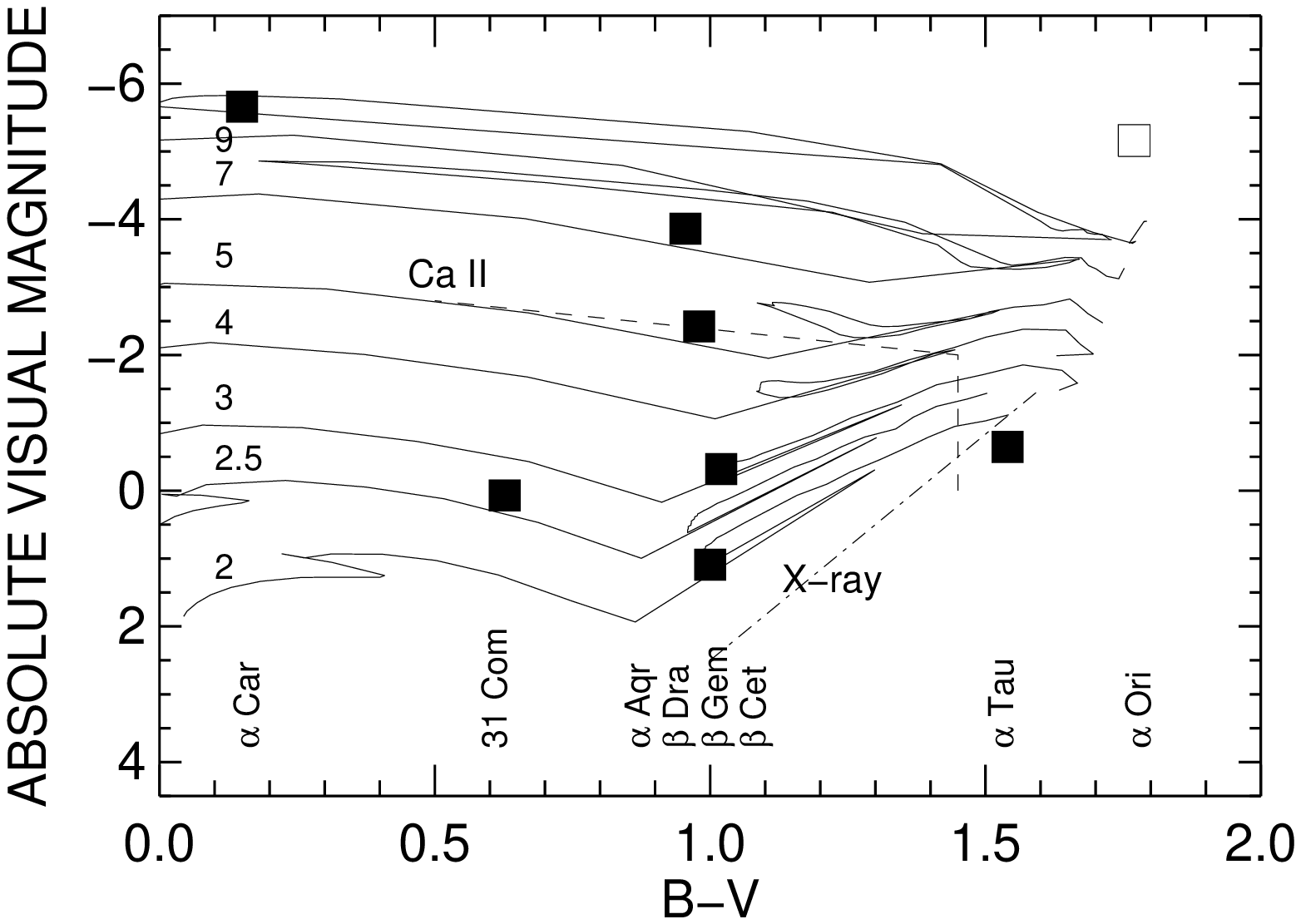}
\caption{Target stars included in the FUSE luminous star survey.  
Evolutionary 
tracks are shown for stars of solar mass 2 - 9 (Schaller \etal\
1992); the locus marking the `disappearance' of X-rays in giant
stars (H\"unsch \& Schr\"oder 1996) is indicated ({\it dot-dash}, 
labeled X-ray).
Stars in the region above and to the right of 
the dashed line (labeled Ca II) exhibit narrow circumstellar 
absorption components believed to be associated with a wind (Reimers 1977).
Filled squares mark the stars having
C~III and O~VI emission; \gal\ Ori ({\it open square}) displays neither 
C~III nor O~VI emission. \label{targets}}
\end{figure}

Eight luminous stars: \gb\ Ceti (HD 4128), \gal\ Ori (HD 39801),
\gal\ Tau (HD 29139), \gal\ Car (HD 45348), \gb\ Gem (HD 62509),
31 Com (HD 111812), \gb\ Dra (HD 159181), \gal\ Aqr (HD 209750) 
were selected by the Cool Stars team
on the \fuse\ satellite in order to obtain far UV spectra of 
objects of various effective temperatures, degree of activity,
and luminosity (see Fig.~\ref{targets}).   Parameters of these stars are given
in Table~\ref{starparams}.  Analysis of these spectra is presented
here.  A  complementary paper on the survey of cool dwarf stars
with \fuse\ is reported by Redfield \etal\ (2002).

\section{Observations and Data Reduction}
The \fuse\ instrument and its calibration are discussed in
Moos \etal\ (2000) and Sahnow \etal\ (2000).
\fuse\ has 4 co-aligned prime focus telescopes that feed light
to four Rowland spectrographs.  Two of the spectrograph gratings
are coated with LiF and two with SiC (Moos \etal\ 2000), enabling 
full (and redundant, in some regions) coverage
of the \fuse\ wavelength range: 905\AA--1179\AA. 
The spectral segments are denoted
by the grating coating and telescope (e.g. LiF1, LiF2 and SiC1, SiC2) and by the detector
(A or B). The monochromatic spectral resolving power of \fuse\ is
20000$\pm$2000
(\fuse\ Observers Guide V4.0) or $\sim$15 km/s. With 
good signal-to-noise in a line profile,
and the oversampling of the FUSE spectrum, the position of a 
spectral line can be determined to about 2 \kms.

All spectra were obtained through the large aperture of \fuse\ 
(denoted LWRS, a 30 arcsecond square);
\gal\ Tau  was also observed through the 4 $\times$ 20 arcsecond square medium 
aperture (MDRS) in order to minimize airglow contamination.
Details of the \fuse\ observations are noted in Table~\ref{exposures}. Spectra 
were reduced with the \fuse\ CalFUSE v2.0.5 pipeline, except 
CalFUSE v2.4 was used for \gal\ Tau in the LWRS aperture.  
To combine exposures, individual subexposures for a single 
telescope/detector combination (i.e. LiF1A, SiC2A etc.)
were co-added after alignment using cross correlation techniques.
Restricted wavelength ranges for the cross correlation 
were chosen for each detector segment so as 
to avoid geocoronal emission  and to perform
the cross-correlation alignment on strong stellar features.
Extractions of spectra obtained only during night time pointings
were made for all targets in addition to the normal procedures of
extracting `both' spectra (both day and night combined).

Individual images were examined to ensure that exposures
with burst characteristics were not included and the star was in
the aperture in all channels.  However it is difficult to
identify placement in the MDRS aperture where the star may be close
to the edge of the aperture (cf. also Redfield \etal\ 2002).  Data
from the large aperture, LWRS are used for flux measurements of all 
targets. Special attention
was given to the SiC channels to verify that the target was
in the aperture. The spacecraft guiding is maintained using 
the LiF1A channel,
and the SiC channels can become misaligned.  In several 
subexposures of \gal~Tau and
\gb~Dra, the C III 977\AA\ emission was not visible
and we eliminated that subexposure from the summations 
(see Table~\ref{exposures} for a summary).

Another potential contaminant is scattered sunlight that can
affect both the fluxes and profiles of the C III (\gla977) and
O VI (\gla1032) transition.  Comparison of day and night
extractions can identify the presence of a solar component.  Scattered
sunlight is present
in the  \gal\ Tau spectrum in the LWRS aperture, 
where the night time extractions
were used for the SiC2A and LiF1A channels (see Table~\ref{exposures}).

\begin{figure}

\includegraphics[angle=90, scale=.8]{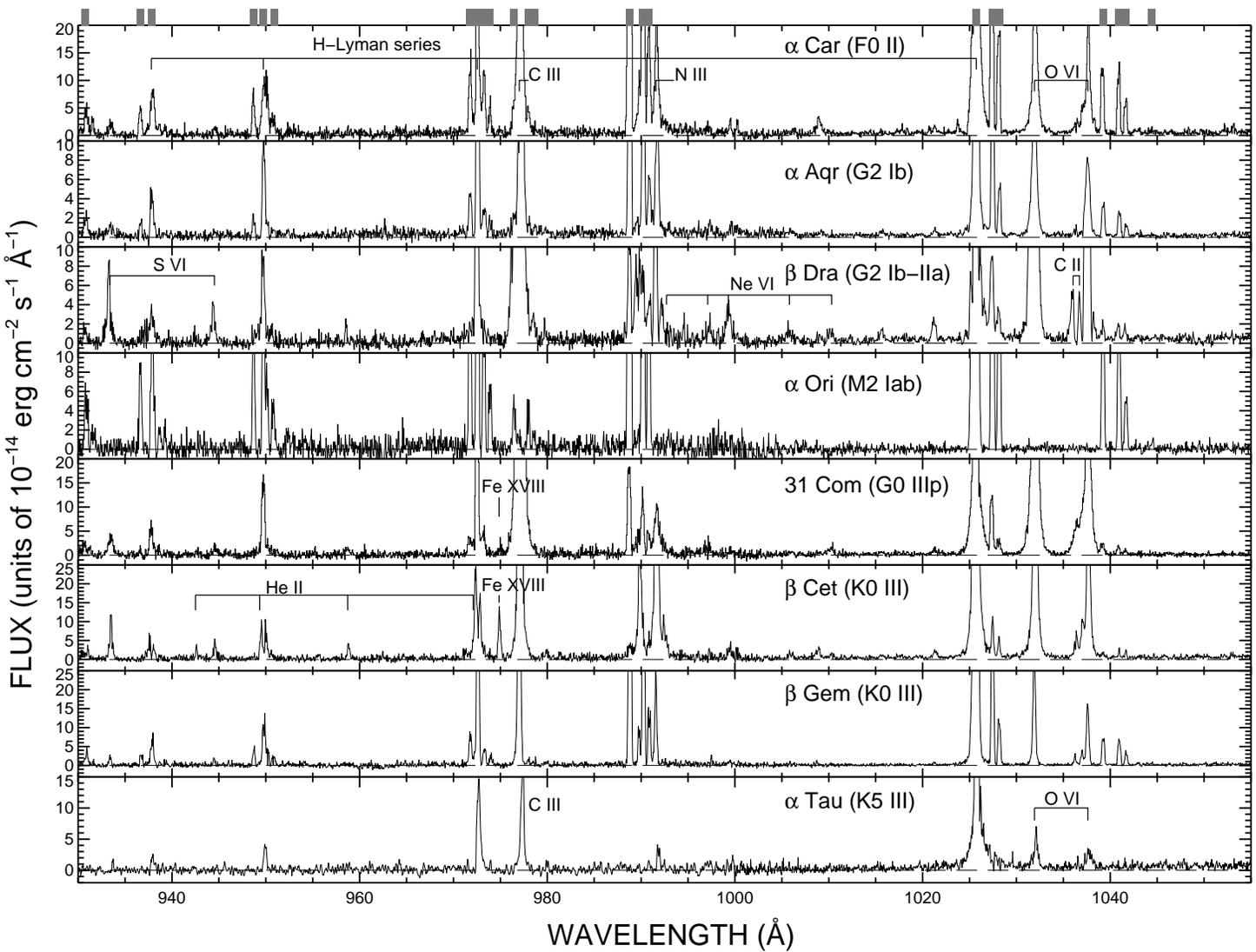}
\caption{FUSE spectra of the sample of luminous cool stars in
the region $\lambda\lambda$ 930--1055  
Spectra from each detector (SiC2A and LiF1A) have been cross-correlated and
summed for all images, then rebinned by a factor of 8 for display.   Prominent
emission features are identified.  Strong airglow lines are frequently
truncated and their positions are 
noted by the hatched area at the top of the figure.\label{spec.short}}
\end{figure}

\begin{figure}
\includegraphics[angle=90, scale=0.8]{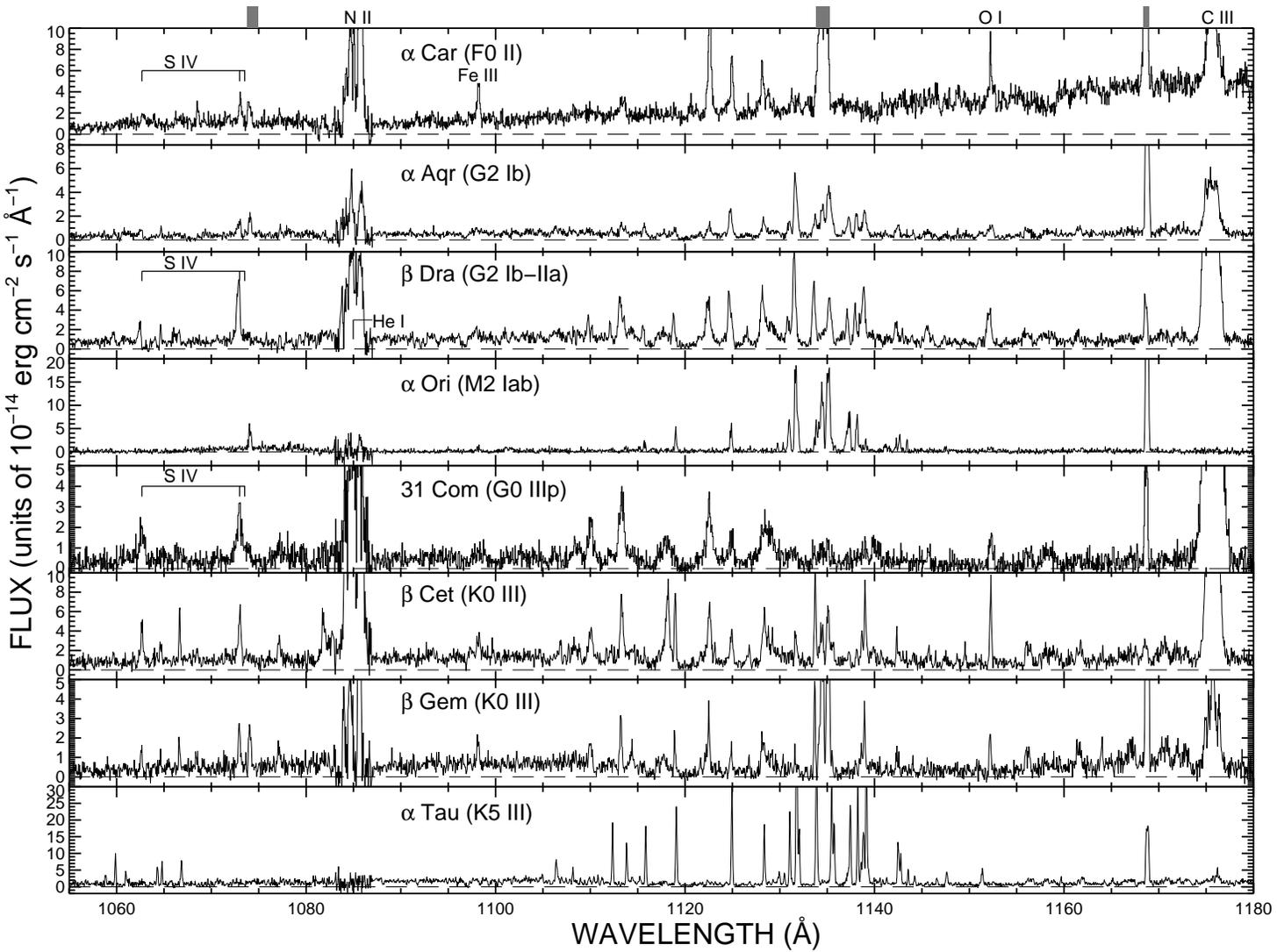}
\caption{FUSE spectra of the sample of the cool star survey in
the range $\lambda\lambda$1055--1180.  Detailed identifications
of the complex region between 1110--1145\AA\ are given
in Fig~\ref{sp.1140}.  Shaded regions at the top of the
figure indicate positions of potential contamination by 
geocoronal emission.\label{spec.long}}

\end{figure}

Short wavelength spectra for the 8 targets are 
shown in Fig.~\ref{spec.short},
and the corresponding long wavelength spectra 
are shown in Fig.~\ref{spec.long}.

\section{Wavelength Scale}
The relative wavelength scale for each detector segment 
of each channel is 
determined by the CalFUSE pipeline and believed to
be accurate to 3--4 pixels ($\sim$ 0.025\AA\ or 8 \kms) over most of the
detectors\footnote{See the FUSE Wavelength Calibration: A FUSE White
Paper at\hfil\break {\it http://fuse.pha.jhu.edu/analysis/calfuse\_wp1.html}}.  However
the absolute velocity offset for each detector must be determined
independently in each detector segment. 
Narrow interstellar absorption lines of C III (\gla 977) and C II (\gla 1036) can
be identified in the spectra from the SiC2A  and LiF1A  channels respectively
where they occur in the profiles of the stellar emission lines.
Ultraviolet spectra from HST (STIS/GHRS) typically contain interstellar
lines of low ionization species (C II, Si III, D I,  O I, and Mg II)
whose velocities are measured; these values are
used to set the absolute wavelength scale of the \fuse\ 
spectra (see Table~\ref{ism}). Two stars, \gal\ Aqr and \gb\ Dra,
show interstellar H$_2$ near \gla 1038, 
and interstellar C II (\gla 1036.34) can be identified
in \gal\ Aqr.  However, in the LiF1A channels, 
4 targets (31 Com, \gb\ Gem,
\gal\ Ori and \gal\ Tau) show no discernible interstellar feature 
and had to be treated differently. For \gb~Gem, several 
chromospheric lines [C II (\gla 1036.3367, short wavelength wing
only), C~II (\gla 1037.0182), S~IV (\gla 1062.621), Si~IV (\gla 1066.6094),
and S~IV (\gla 1072.9558)] were fit to Gaussian curves and
assigned a photospheric velocity.  The average offset for
\gb\ Gem was determined to be  0.0351\AA $\pm$0.008.
The star 31 Com is more difficult because only 2 weak stellar S IV lines are
available (\gla 1062.6166 and \gla 1072.9954) that give offsets
of $-$0.0571\AA\ and $-$0.0099\AA\ respectively -- an uncomfortable
spread of 13 \kms. Having no other alternative, we take
the average offset ($-$0.0335\AA) to set the absolute wavelength scale. 
For \gal\ Ori and \gal\ Tau, the CalFUSE 2.0.5 wavelength
scale is adopted here. 

\fuse\ channels at the 
longest wavelength (\gla\gla1100--1180) generally do not contain 
observable interstellar lines in
the spectra of cool stars.  In this case we resort to HST spectra 
obtained wtih the Space Telescope Imaging Spectrograph ({\it STIS}) or
the
Goddard High Resolution Spectrograph (GHRS).  For the
LiF 2A (and Lif1B) channel we adopt the HST/STIS/GHRS
absolute wavelength offset by forcing agreement between the 
\hst\  and \fuse\ wavelengths using C III 1176\AA\ when observed
in both spectra.  Otherwise, we have forced the velocity 
of the ions observed in  LiF2A (Si~III, \gla 1113.23
and Si~IV, \gla 1122.49) to match the same ion,  
Si~III (\gla 1206) and Si~IV (\gla 1394, \gla 1401)   
observed with HST.

The velocity offsets necessary for alignment with previously
observed interstellar 
line velocities are usually small. For  6 targets of 
this program (not including \gal\ Tau and \gal\ Ori), the
corrections for the C III line (SiC 2A channel)
ranged in absolute value from 4.4 to 29 \kms\ with
an average of 11.5 \kms.  The LiF2A channel for the
same 6 targets gave offsets ranging from $-$0.0335\AA\ ($-$9.7 \kms)
to +0.06873\AA\ ($+$17.5 \kms) with an average of 0.0245\AA\ ($+$6.25 \kms).

Comparison of the wavelength offsets obtained in
this way were made with  wavelengths of unblended O I airglow lines observed
in the spectrum for several cases.  Airglow lines typically vary in
wavelength by 5 to 18 \kms\ from the offset determined by the
interstellar features, so they can give a crude estimate of the
absolute  scale, but only to $\sim \pm$10 \kms. For \gb\ Gem, the O I airglow
gives an offset for SiC2A of +17.5 \kms\ as compared to the
interstellar offset of $+$15.0 \kms. In the LiF2A channel, the
airglow values vary from $+$5 to $+$18 \kms\  different from the
offsets determined by interstellar lines.  Although lack of
precision in the wavelength scale exists, this does not
materially affect the line identifications and fluxes, or 
conclusions drawn from
the analysis of line profiles.

\section{Line Identifications}

Line identifications were made by comparison to the solar
spectrum (Curdt \etal\ 2001), and to other cool luminous
stars observed by \fuse\ (Young \etal\ 2001; Ake \etal\ 2000).
The highest ionization lines (\ion{Fe}{18} and \ion{Fe}{19}) are
discussed separately below.  The strongest species are marked
in Fig. 2 and 3. Every star, except for \gal\ Ori exhibits
emission from C III and O VI, indicating that temperatures
at least as high as 3$\times$ 10$^5$K are present assuming 
a collision-dominated thermal plasma.
Emission from the Earth's atmosphere (``airglow'') is
identified using several spectra with long integrations
on the sky.\setcounter{footnote}{2}\footnote{We 
extracted a sky exposure of $\sim$58 ks from the
LWRS from Observation P1100301; another airglow 
spectrum can be found on the \fuse\ 
website, {\it http://fuse.pha.jhu.edu/analysis/airglow/airglow.html}.}

Fluxes were extracted (Table~\ref{tblflux}) for the strongest lines of C III
(\gla977 and \gla1176) and O VI (\gla 1032) by integrating directly over
the line profiles.  These fluxes agree with previous ORFEUS
measurements (Dupree \& Brickhouse 1998) 
of  \gal\ Aqr and \gb\ Dra to within 6\% on average; 
an exception is \gla 1176 which is stronger
in the \fuse\ spectrum of \gal\ Aqr  by a factor of 2.5.  
Such a large discrepancy is unexpected.  Because the 
\fuse\ spectra for these 2 stars were taken through the large
aperture, they are not subject to flux loss as found in
some medium aperture spectra (Redfield \etal\ 2002).  The supergiant
\gal\ Aqr exhibits periodic chromospheric variability (Rao \etal\
1993), and variability in the Mg II flux (Brown \etal\ 1996), which 
may account in part for the discrepancy.
Cool dwarf stars studied with \fuse\ (Redfield \etal\ 2002) also
show discrepancies, on average of 20\%, when compared to fluxes 
from ORFEUS. 

\begin{figure}
\includegraphics[angle=0.,scale=0.9]{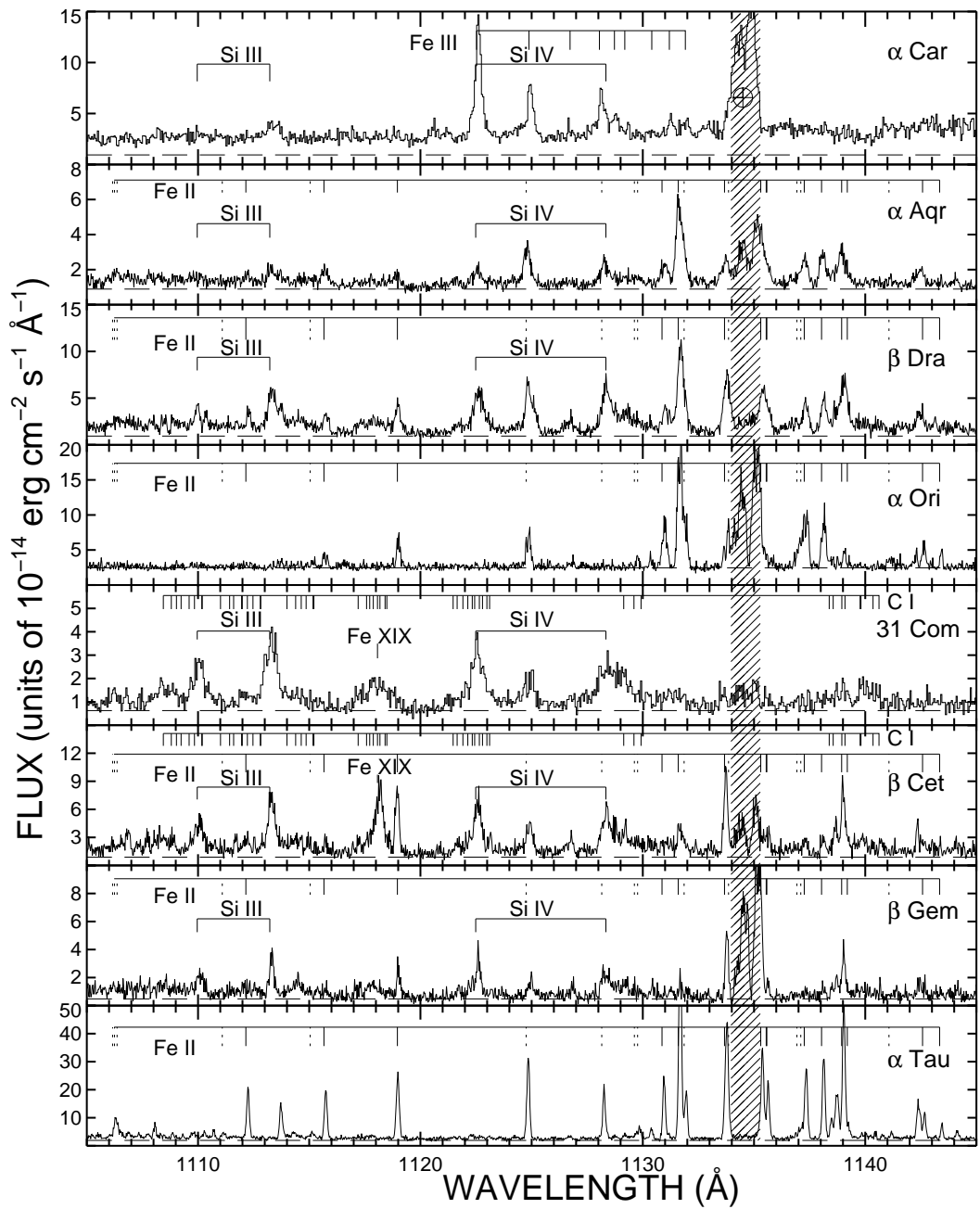}
\caption{The region \gla \gla 1110--1145 showing the presence
of fluorescent Fe II emission in many of the coolest targets.  The
hatched area marks the position of N I airglow emission. The location
of Fe II emission  marked by solid ({\it broken}) lines indicates
those features pumped by radiation within ({\it larger than}) 1.8\AA\ 
of the H Lyman-\gal\ core, and might be expected to be strong ({\it weak}).\label{sp.1140}}

\end{figure}

Luminous cool stars show a distinct pattern of narrow
emission lines near \gla\gla1130--1140  first noted in the ORFEUS spectrum
of \gal\ TrA and ascribed to low-ionization states most probably
fluoresced in the extended cool atmosphere (Dupree \& Brickhouse
1998).  It required the
higher spectral resolution of \fuse\ to identify many of the 
emission lines as Fe~II (Harper \etal\ 2001) that result from
fluorescent decay of levels  pumped
by Lyman-\gal\ (cf. Hartman \& Johansson 2000).  The targets
in this survey also show many of the same narrow lines
(see Fig~\ref{sp.1140}).

Details of the Fe II line strengths are puzzling. Alpha Tau displays a 
strong Fe II spectrum where it appears           that
Fe II lines pumped by radiation close to the Ly-\gal\ core ($<$1.8\AA)
are present and strong, and lines pumped by more distant 
wavelengths are absent or weaker as
suggested by Harper \etal\ (2001) based on a spectrum of 
\gal\ TrA.
The strong signature transitions of Fe II between 1131--1139\AA\
support this conjecture.  However the Fe~II  lines do not have similar relative
strengths in  other targets.  Whereas 1131.594\AA\ dominates in
the spectrum of \gal~Tau, \gal~Ori, \gb~Dra, and \gal~Aqr, another
transition of Fe II at 1138.941\AA\ dominates in \gb~Gem and \gb~Cet, 
appears in \gal~Tau, \gb~Dra, and \gal~Aqr, but
is very weak or absent in \gal~Ori.  Moreover, 
the supergiant \gal~Ori has a  more extensive atmosphere  
than the giant \gal~Tau, which would appear to enhance 
fluorescent processes, yet  
the fluoresced lines appear weaker in \gal\ Ori than in \gal\ Tau. 
Because the presence of the fluoresced lines depends on the
intrinsic stellar Ly--\gal\ profile shape, its flux, and
the detailed atmospheric dynamics to enable the process, 
models specific to each star need to be constructed to
interpret these spectra. 

\begin{figure}
\begin{center}
\includegraphics[scale=0.9]{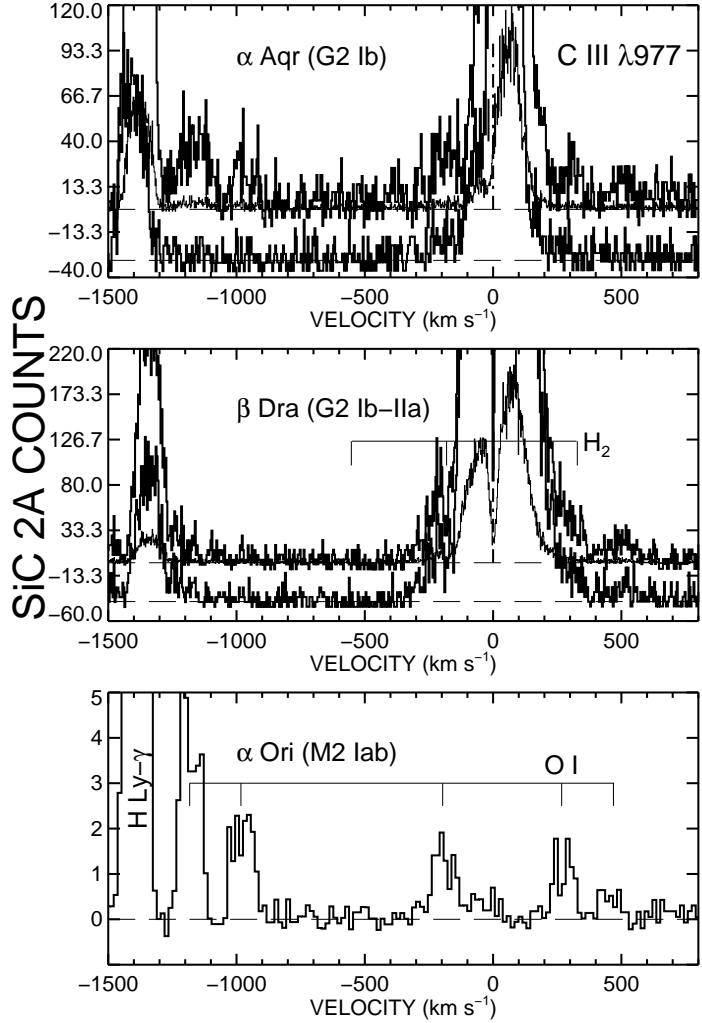}
\caption{The extremely low background of the \fuse\ detectors
allows identification of weak features, here in the
wings of C III, \gla 977.  Total ({\it upper curve})
and night only ({\it lower curve, offset}) extractions are shown for
\gal\ Aqr and \gb\ Dra. These rebinned spectra are shown
at 10 times the original count level.  The original
profile is also displayed ({\it thin line}).. Airglow lines of H-Ly$\gamma$ are present
in all spectra.  The airglow  due to O I is present
in \gal\ Ori, \gal\ Aqr, and very weak, if not absent in the
\gb\ Dra spectrum. 
Emission near $+$500 \kms\ in the 
spectrum of  \gb\ Dra is present in both total and night
extractions demonstrating the ability of \fuse\ to detect such weak
features. It appears likely that this emission results from O I 
(\gla 978.624) that is fluoresced by the stellar C III line itself via
O I transitions sharing the same upper level, $2p^3 5s\ ^3S_1$. 
$H_2$ may also contribute to the absorption feature near $-$200~\kms\ 
in \gb\
Dra; the location of three 
$H_2$ transitions are marked as synthesized by {\it h2tools}. \label{c3.wings}}

\end{center}

\end{figure}

It is worth noting that the low background count rate of the
\fuse\ detectors enables identification of weak emission features.
As shown in Fig.~\ref{c3.wings}, longward of the C III, \gla977
emission, the spectrum of \gb\ Dra exhibits, no airglow,
but an O I line at \gla977.62 that 
is fluoresced most probably by the C III transition itself
through the stellar resonance O~I transition at \gla976.45 with the same 
upper level ($5s\ ^3S_1$), 
overlapping the broad C III profile.

\section{Coronal Lines}
Two  targets, \gb\ Cet and 31 Com contain emission
from high temperature coronal species: Fe XVIII and Fe XIX.  
This is not surprising
since these very same ions have been identified in EUVE
spectra of these stars  (Sanz-Forcada \etal\ 2002).  Similar
transitions were found in the \fuse\ spectrum of
Capella (Young \etal\ 2001) and other targets discussed by
Redfield \etal\ (2003) including \gb~Cet and 31~Com.
Coronal species are present in the near ultraviolet
region covered by HST; these include Fe~XII and Fe~XXI
(Jordan \etal\ 2001; Ayres \etal\ 2003).

\subsection{\gb\ Cet}

\gb\ Cet shows the highest excitation lines in the \fuse\  
spectral region, namely Fe XVIII (\gla974.86, 
$2s^22p^5\ ^2P_{3/2}-2s^22p^5\ ^2P_{1/2}$) and Fe XIX 
(\gla 1118.07, $2s^22p^4\ ^3P_2-2s^22p^4\ ^3P_1$ )
arising from transitions within the ground configurations
of the atom.    The observed  wavelength of the
Fe XVIII, \gla974.85 transition (corrected for the +12.3 \kms\ 
radial velocity of the star)  agrees with the laboratory
wavelength to 0.015\AA\  and confirms that the feature
corresponds to the photospheric velocity. The 
FWHM equals 0.29$\pm$0.02\AA\ which is comparable to
the thermal broadening expected in a plasma at T$=$10$^{6.8}$K. 
The line flux is measured to be 3.6$\pm$0.4 $\times$10$^{-14}$ \ecs.
These parameters, here measured from the photon (counts)  spectrum 
confirm the values in Redfield \etal\ (2003).
The Fe XIX transition appears blended
with a broad C I multiplet that occurs from  1117.2 to 1118.5.
The blend was deconvolved into a broad (FWHM= 1.4\AA) 
and narrow (FWHM=0.33\AA) component.
Line center, corrected for the stellar radial velocity, and 
corrected by using the Si III transition at \gla1113.228 as
a fiducial reference, occurs
at 1118.081\AA\ in agreement within 0.01\AA\ with the laboratory
value.  The expected
fluxes of Fe~XVIII and Fe~XIX from \gb~Ceti were predicted by using 
atomic emissivities from 
CHIANTI/APEC (Dere \etal\ 2001; Smith \etal\ 2001) and an 
emission measure distribution
from iron lines measured in the EUVE spectrum in 2000 (Sanz-Forcada 
\etal\ 2002); the observed fluxes in \fuse\ spectra are 
stronger than the predictions by a factor of
$\sim$1.7 for \gla974, and by a factor of 1.6 for \gla 1118. 
The ratio of the fluxes of Fe~XVIII/Fe~XIX is predicted to be
\gla974/\gla1118=1.8, as compared to the measured value
of 2.02, an amount that is within 12\% of the prediction. 

The agreement of the flux values, within a factor of 2 is considered acceptable,
based on the uncertainties in atomic parameters, density effects,
calibration errors, the interstellar absorption correction,  and
possible variations in the source itself.
Because \gb\ Cet
became much more active in August 2000 as compared to the
earlier EUVE observation in 1994 and displayed frequent flaring
activity not found earlier (Ayres \etal\ 2001b; 
Sanz-Forcada \etal\ 2003),   
this activity may have continued
through the \fuse\ observations in December 2000 although we
have no direct evidence of  continued activity.  Certainly, a
variation of a factor of two in highly ionized species
is not surprising as has been noted earlier in Capella (Dupree \&
Brickhouse 1995).
And it is also possible there are additional cascade contribution from higher
levels of Fe XVIII and Fe XIX that are not included  
in the CHIANTI/APEC emissivities leading to an underestimate 
of the predicted value.

\subsection{31 Com}

It appears likely that Fe XVIII and Fe XIX are also present in
the spectrum of 31 Com.  31 Com has a
high surface flux of O VI and an emission measure distribution 
derived from EUVE spectra that mimics that of \gb\ Ceti (Sanz-Forcada \etal\
2002), except the coronal enhancement occurs at slightly higher temperatures.
Using the EUVE emission measure distribution, the
flux of Fe~XVIII (\gla 974) is predicted to be
1.8$\times$10$^{-15}$ \ecs\  and Fe~XIX (\gla 1118) is estimated as 
1.63$\times$10$^{-15}$ \ecs.  
Although weak\footnote{In addition to intrinsic weakness, the Fe~
XVIII is affected by the C III 977\AA\ broad wing which contributes to elevate
the background.  Taking the binned spectrum, we fit multiple Gaussian 
profiles to the C III wing,
Fe XVIII, and nearby airglow lines as well as the linear continuum.
The FWHM of the Fe XVIII emission was determined as $\sim$0.40\AA,
and so the summing of the spectrum was made to $\sim$1.65$\sigma$ which
should include $\sim$90\% of the line emission for the Fe XVIII feature.} 
Fe XVIII is observed in the SiC2A channel with a flux of  
4.6$\times$10$^{-15}$ \ecs, a factor 2.5 times larger than predicted.
The Fe~XIX flux 
is difficult to measure because it is located in a complex
of C I emission.  To estimate the flux, we scaled the expected C I strength in
the blended multiplet at \gla 1118 in 31 Com from the adjacent 
C I multiplet near \gla 1115 using the ratio measured in the quiet sun
spectrum (Curdt et al. 2001).  The cell center and network ratio
differ only by 7\%.  Fe XIX observed in the LiF2A channel  equals
6.62$\times$10$^{-15}$ \ecs.   As in the case of \gb\ Cet, the observed
\fuse\ values are larger than predicted, in this case by a factor of 4.
The flux ratio Fe~XVIII/Fe~XIX is predicted to be 1.1 as compared to
the observed value of 0.7, representing a difference of a factor of 
1.6.  Such a discrepancy is not surprising considering the flux
extraction procedure in addition to other uncertainties noted above.

\section{Lyman Series Emission}

\begin{figure}
\includegraphics[scale=0.8]{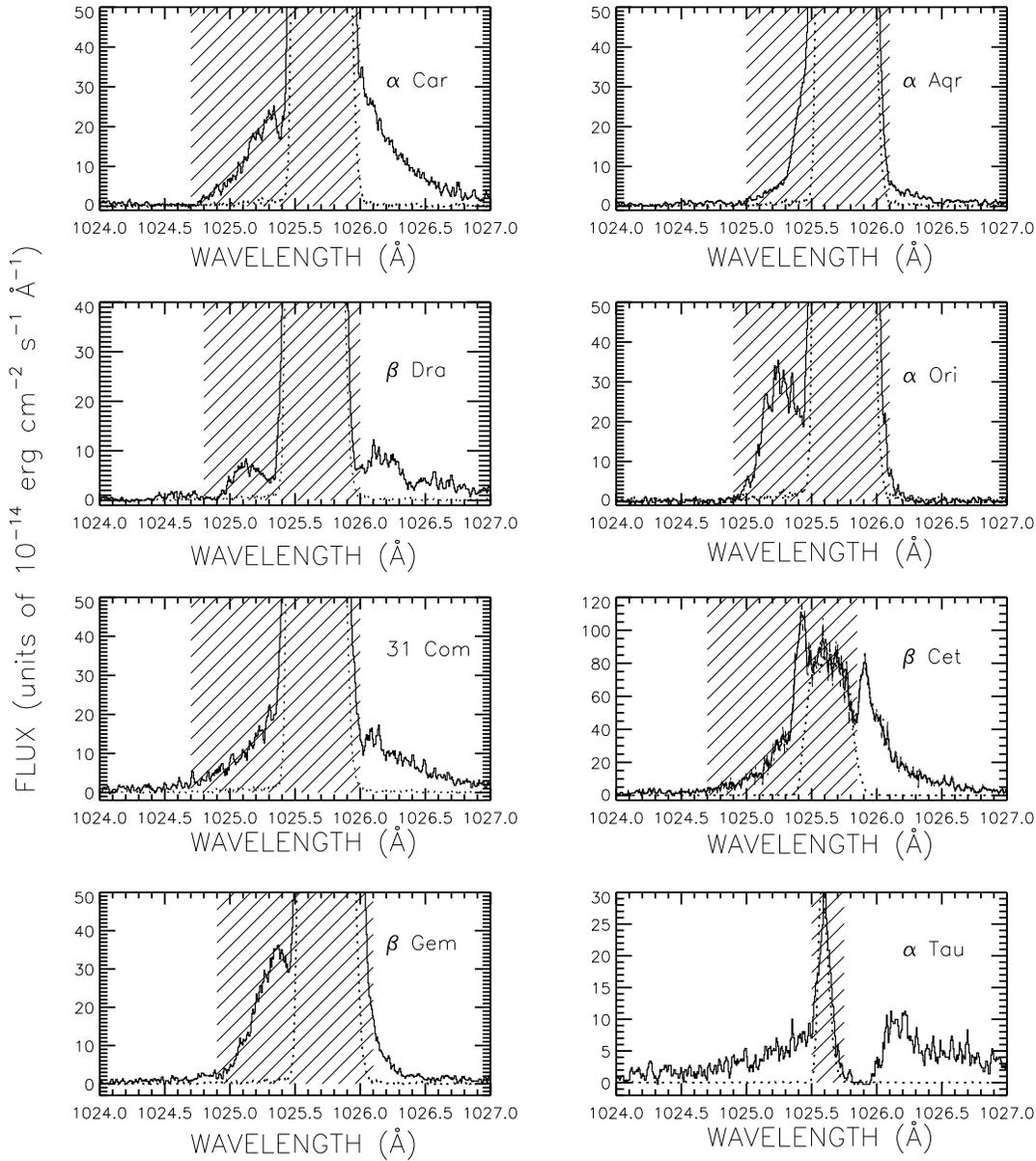}

\caption{The Ly-\gb\ profile in the target stars from
the total (day $+$ night) exposures in LiF1A. A scaled
spectrum of the Ly-\gb\ airglow profile taken in 
August 1999 through  the LWRS or MDRS, as 
appropriate, is shown in each panel ({\it broken line}).  The
hatched area indicates where spectra are most likely
affected by airglow and detector walk caused by gain sag 
({\it see text}). Excess
emission from the stars on the long wavelength side of
the Ly-\gb\  profile,  appears in all targets except 
\gal\ Ori.  Note the narrow appearance of
the Ly-\gb\ airglow through the  medium aperture (MDRS) in the \gal\ Tau 
spectrum. With this aperture, detector walk has not occurred. The
night-only extraction shows little difference on the long
wavelength wing from the total data, but is noisier because of the
shorter exposure. Data for
\gb\ Cet are only nighttime exposures. \label{sp.lyman}}

\end{figure}

All of the stars display a Lyman-\gb\ emission feature at
\gla1025.  This is the strongest airglow line in the \fuse\ spectral range, 
and is present both in day and night spectra.  Unfortunately
the strong airglow has caused a drop in the gain of the detectors
in this wavelength region when using the large aperture.  Because
the reported x-positions of the photons arriving  on the detector are
a function of the gain, the drop in gain (`gain sag') causes lower gain
events to be incorrectly positioned ({\it cf.}  The 
\fuse\ Instrument and Data Handbook, V. 2.1, Sec. 9.1.12).  The tendency 
for photons to be moved to shorter wavelengths on
the LiF1A detector, causes emission features to
appear that can mimic actual stellar emission.  The nature
of the observed emission is revealed by  inspection of 
the pulse height distribution of the feature in the raw data,
and extraction of  the spectrum with CalFUSE using various
levels of the pulse height screening parameter. 
Emission features on the short wavelength side of Ly-\gb\  
in Fig.~\ref{sp.lyman} become weaker when the pulse height
threshold is raised.  The \fuse\ project remedies the gain sag
problem periodically, but it is usually present at some level
making suspect the emission features that lie shortward of Ly-$\beta$.

In most of our targets there is excess emission to the
{\it long} wavelength side of Lyman-\gb\ (Fig.~\ref{sp.lyman}).
To identify the stellar emission in this complex feature, 
the profiles are compared to \fuse\ reference airglow spectra. An extended
exposure on the ``sky'' was taken  in August of 1999 yielding spectra
through all \fuse\  apertures with no target in
the field. Because  this date was very early in the \fuse\ mission,
the profiles are not affected by gain changes in the detectors. 
These spectra are available on the \fuse\ website,
and the LiF1A spectra were scaled to the observed airglow profile
in the target stars.  Spectra from the LWRS were used for
all stars, except for \gal\ Tau where a MDRS spectrum was
substituted.   The profile of the Lyman-\gb\ 
airglow emission is the same for both day and night extractions,
although the flux level is lower in spectra obtained at night. 
While absorption by interstellar deuterium is expected
at \gla1025.443, airglow in the large aperture contaminates
this region.  Spectra of \gal\ Tau, taken through the medium
aperture, do not have sufficient signal to detect  D~I
absorption.

All of the targets except \gal~Ori have excess emission
on the long wavelength side (Fig. 6) which arises from H~I in the stellar 
chromosphere. Emission is expected and likely to be 
self-reversed because  Lyman-\gb\ is
an optically thick chromospheric line.  Additionally, motion in the
atmosphere can create asymmetries in the profiles, and 
absorption by interstellar hydrogen can substantially change 
their appearance.  It is not possible to draw conclusions
about the intrinsic stellar line flux or shapes because of airglow
contamination and instrumental effects.   The H Ly-\gb\ profile
of \gal\ Tau, because it was taken through the MDRS, reducing the
airglow contamination, comes closest to sampling the
stellar profile but the interstellar absorption at $-$30 \kms\ coupled
with the stellar radial velocity of $+$54 \kms\ affects
the central reversal. In spite of problems with the line core and blue
wings, the extent of the H Ly-\gb\ long wavelength emission wings indicates
the width of the H-Ly\gal\ profiles.  The 
H Ly-\gb\ stellar emission wings on the long wavelength
side span a region 0.5 -- 1.0\AA. 
Because the H Ly-\gal\ line width is about a factor of 1.4 broader
than the H Ly-\gb\ line in the Sun (Lemaire \etal\ 2002), 
it appears that sufficient flux is present in the stellar H Ly-\gal\ wings to pump 
Fe~II and cause fluorescence observed in the spectra shown
in Fig.~\ref{sp.1140}.  The fact that \gal\ Ori does not show any
stellar emission in Ly-\gb\ may provide the explanation for the
weakness of the fluoresced Fe II emission line near \gla1135 noted
earlier in Section 4.

\section{Time Variation: Beta Ceti}

\begin{figure}
\centerline{\epsfxsize=10cm\epsfbox{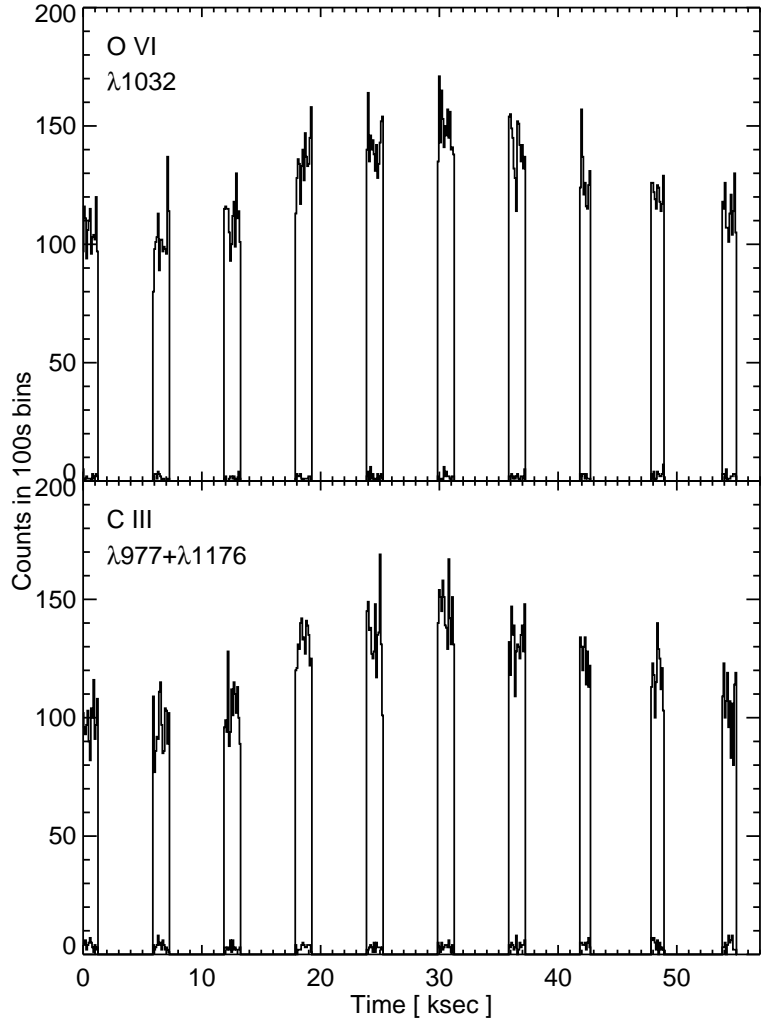}}
\caption{Light curves for \ion{O}{6} (upper panel) and \ion{C}{3}
(lower panel) during the observation of  $\beta$ Ceti. The data have
been placed into 100~s time bins as described in the text. In each panel the
light curves for the background level are also shown, demonstrating
the
low background levels of \fuse. \label{o6-c3-lc}}

\end{figure}

One star in our sample, Beta Ceti, showed substantial flux 
variation during the exposure. 
Light curves (Fig.~\ref{o6-c3-lc}) were created by considering the raw,
time-tag \fuse\ data. Each of the 10 $\beta$ Ceti exposures was
combined into a single time-tag file using the TTAG\_COMBINE routine
in the \fuse\ software. The detector image from this combined data-set
was inspected and an area bounding the emission line
selected. Another area of the same size lying either above or below
the spectrum was also selected to estimate the level of the detector
background. 

For each area, the number of photons arriving in  100~s time bins
was determined throughout the observation. 
The \ion{O}{6}  light curve was created from the
LiF1A and LiF2B \gla1032 lines, while the \ion{C}{3} light curve was
created from the \gla977 lines in the SiC1B and SiC2A channels summed
with the \gla1176 lines from the LiF1B and LiF2A channels.

\begin{figure}[h]

\plottwo{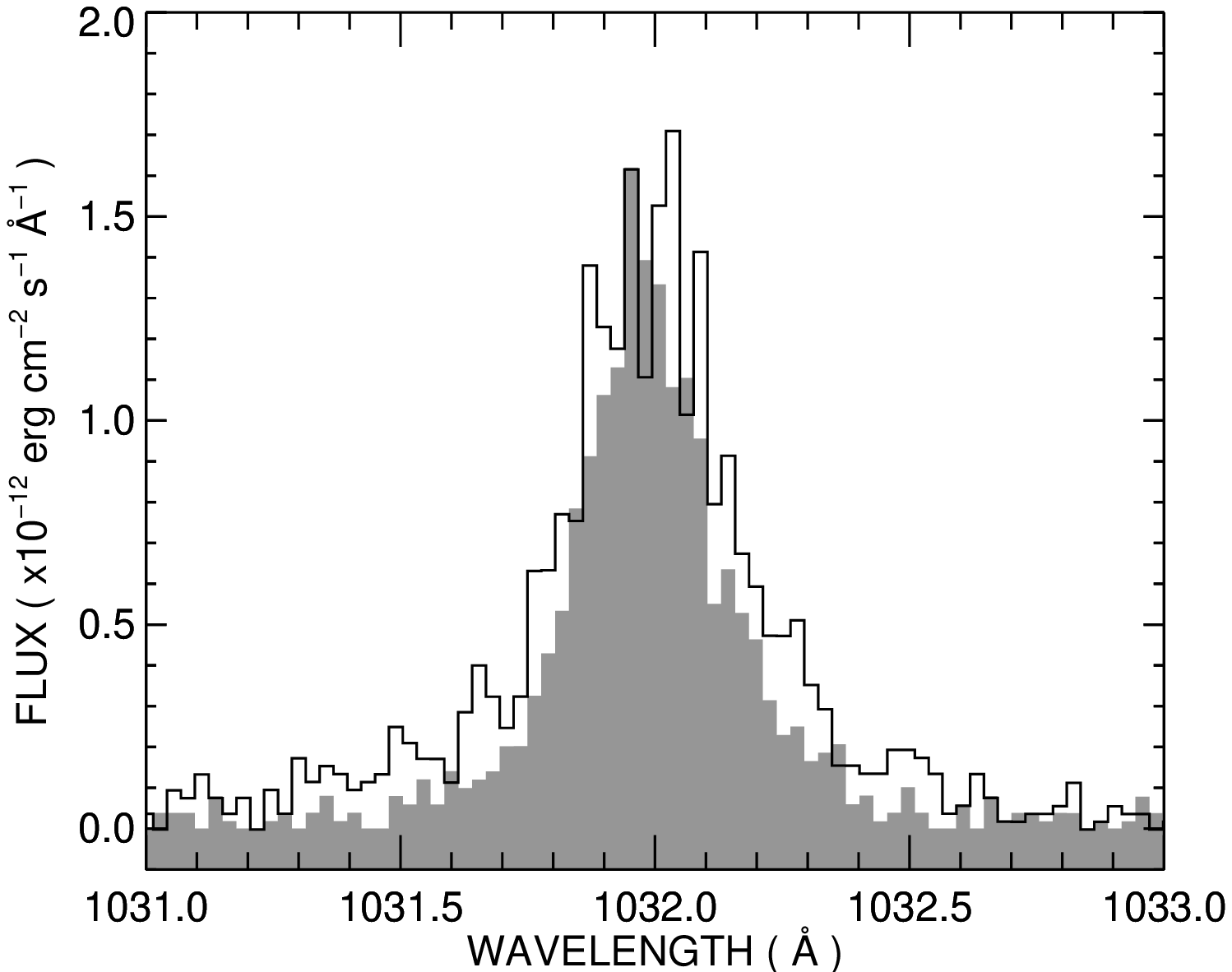}{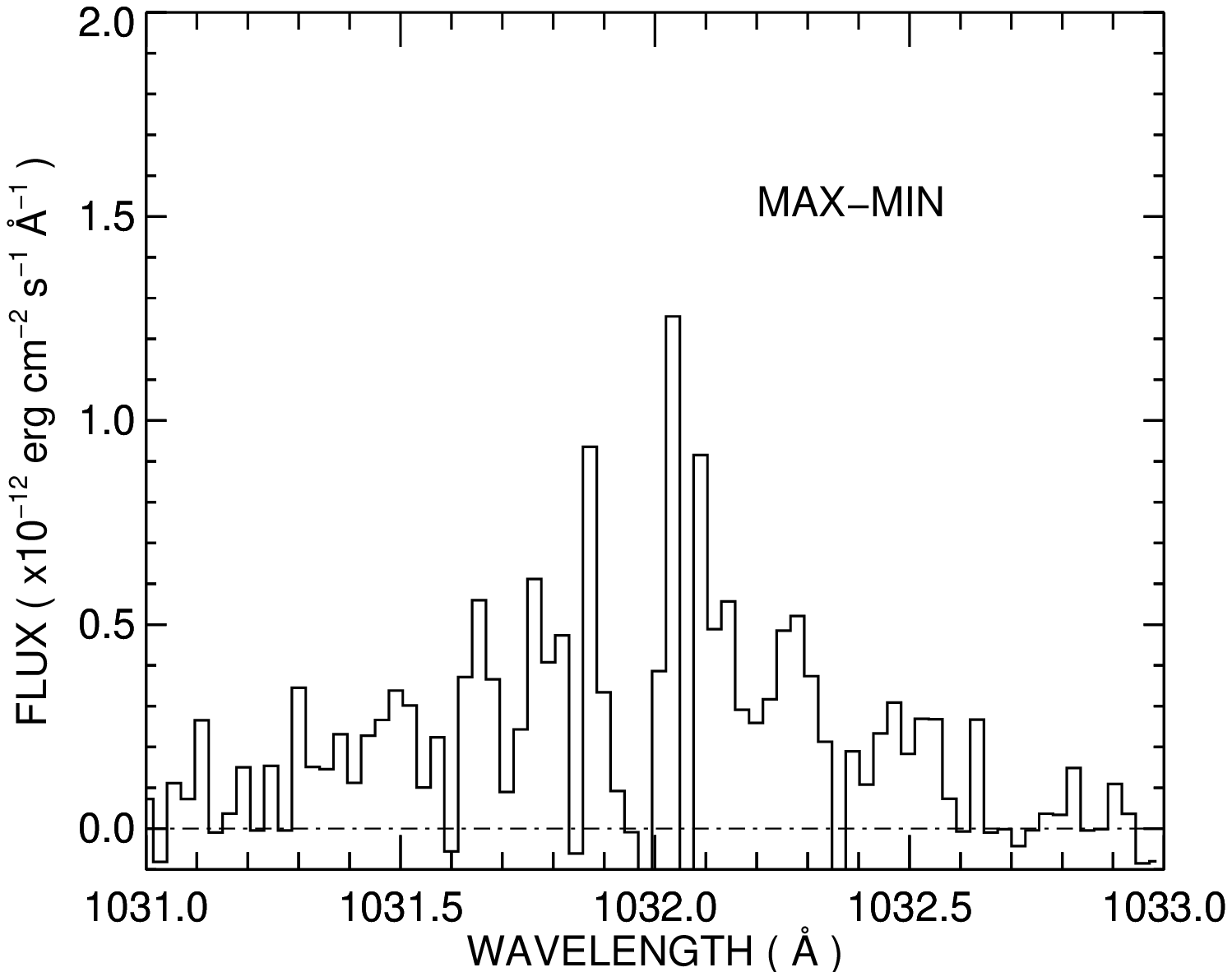}
\caption{{\it Left:} Comparison of \ion{O}{6} \gla1032 profiles from the 1st 
({\it shaded}) and
6th $\beta$ Ceti exposures, corresponding to the minimum and maximum of
the light curve, respectively. It can be seen that the increase in flux
is due to a broadening of the line profile. {\it Right:} The 
difference between the O VI profiles at maximum 
and minimum light illustrating the appearance of emission
about $\pm$0.2\AA\ ($\sim\pm$60 \kms) from line center while emission at
line
center remains effectively constant.\label{bceti-o6}
}

\end{figure}

The \ion{C}{3} and \ion{O}{6} light curves show the same features, namely a
rise in the fluxes of the lines by around 50\%  
during the observation,
followed by a fall to the original flux level at the beginning of the
exposure. The rise and fall times are comparable at $\sim$20ks each.
The increase in flux of the \gla1032 line is simply due to a broadening of
the line profile as illustrated in Fig.~\ref{bceti-o6}.  The flux
at the center of the line remains constant, and the added emission
arises at both high and low velocities from line center (Fig.~\ref{bceti-o6}).
Three other flaring events observed in O~VI with \fuse\ exhibited different
profiles.  AB Dor had a redshifted emission component in O~VI that extended
to 600 \kms (Ake \etal\ 2000).  During  flares from  AU Mic  
an {\it enhanced} core appeared in O~VI, in 
addition to broad wings (in one flare)
or red shifted emission (in another flaring event) (Redfield \etal\ 2002).
Thus, this \gb\ Cet event remains unique with its symmetric broadening
and
constant core.

Coronal emission in \gb\ Cet measured with \euve, exhibited 
flaring events during August 2000 (Ayres \etal\ 2001b;
Sanz-Forcada \etal\ 2003), however they lasted longer than one day. 
\gb\ Ceti is a slow rotator ($vsin i$= 4 \kms; Fekel 1997) so 
the O~VI enhancement does not appear related
to the passage of active regions across the disk, and most likely
represents a long chromospheric-transition region flaring episode.

Many  stars exhibit rapid flux increases in transition 
region lines Si IV and C IV.
However these flares 
typically have rise times, on the order of a few minutes
or less [cf. the dwarf stars, AD Leo (Bookbinder \etal\ 1992),
AB Dor (G\'omez de Castro 2002), and AU Mic (Robinson \etal\ 2001)].
The active dwarf binary HR 1099  showed a 
rise time of about 1.5 hours in one 
event (Ayres \etal\ 2001a), but no events 
comparable to the 5.5 hr rise observed here. A RS CVn-type
binary, $\lambda$ And did undergo 
ultraviolet flaring (Baliunas \etal\ 1984) in
an event that lasted for about 5 hours.  Although most 
normal single giants have not shown transition region flaring,
the bright giant $\lambda$ Vel (K4Ib-II) and \gb\ Ceti 
have exhibited coronal flare episodes lasting from
minutes to days (Ayres \etal\ 1999, 2001b; Sanz-Forcada et al. 2002). 
It is plausible that the enhancement of Beta Cet in
the transition region lines corresponds to such a coronal event
detected  in these other stars.

A few transition region line profiles have been measured
during  stellar flares but no consistent pattern emerges. The
flare star AD Leo showed a substantial (up to 1800 \kms) redshift
in its C IV $\lambda$1550 emission during a flare (Bookbinder \etal\ 1992). 
In HR1099, broad and narrow components 
of transition region lines 
of Si III, Si IV, C IV, and N V  remained present,
but the flux of one or the other component increased in flares
(Ayres \etal\ 2001a). 
In AU Mic, a single broad line of Si IV appeared that
alternately became shifted towards longer and shorter
wavelengths during a flare (Linsky and Wood 1994).
Another flare in AU Mic showed no change in the Si III line profile
but simply a flux enhancement (Robinson \etal\ 1992).
Line profiles during flares in the active rapidly rotating 
dwarf, AB Dor are not always the same, but frequently 
show redshifts (G\`omez de Castro 2002, Ake \etal\ 2000),
although rapid  broadening of the C IV lines to 
several hundred \kms\ is  observed in the strongest
flares.  The broadening of the Beta Ceti profile is within
the range of diverse profiles found in other stars during
flares, but the rise time appears anomalously long.

\section{Density Diagnostics}

\begin{figure}
\epsscale{0.4}
\plotone{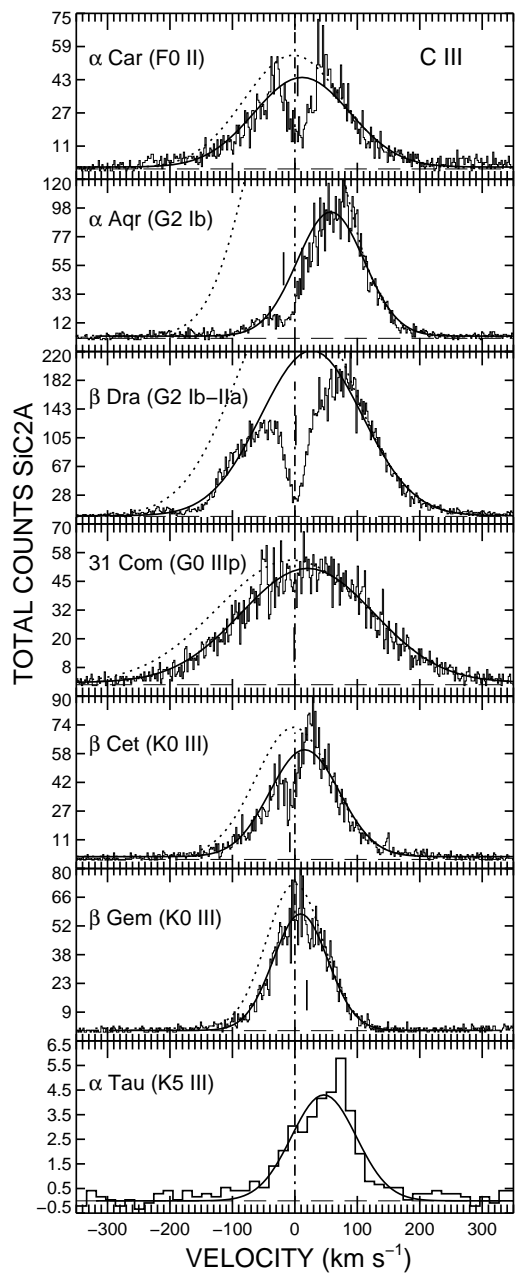}
\caption{The C III, \gla 977.020 emission in the target stars as
measured in the SiC2A channel. The zero point of the photospheric
velocity scale is indicated by a dot-dashed line. Single
Gaussians have been fit to the profiles ({\it solid line}) and to the 
positive velocity side
of the profiles ({\it broken line}).  A short solid line marks the position 
of the interstellar C III absorption.
Line profiles are not rebinned or 
smoothed  except for \gal\ Tau which
is rebinned by 8 pixels.  C~III emission is detected in 
all targets, and appears not to be symmetric in most, but exhibits 
absorption at negative velocities. The \fuse\ wavelength scale was
used for \gal\ Tau. \label{c3.977}}

\end{figure}

\begin{figure}
\plotone{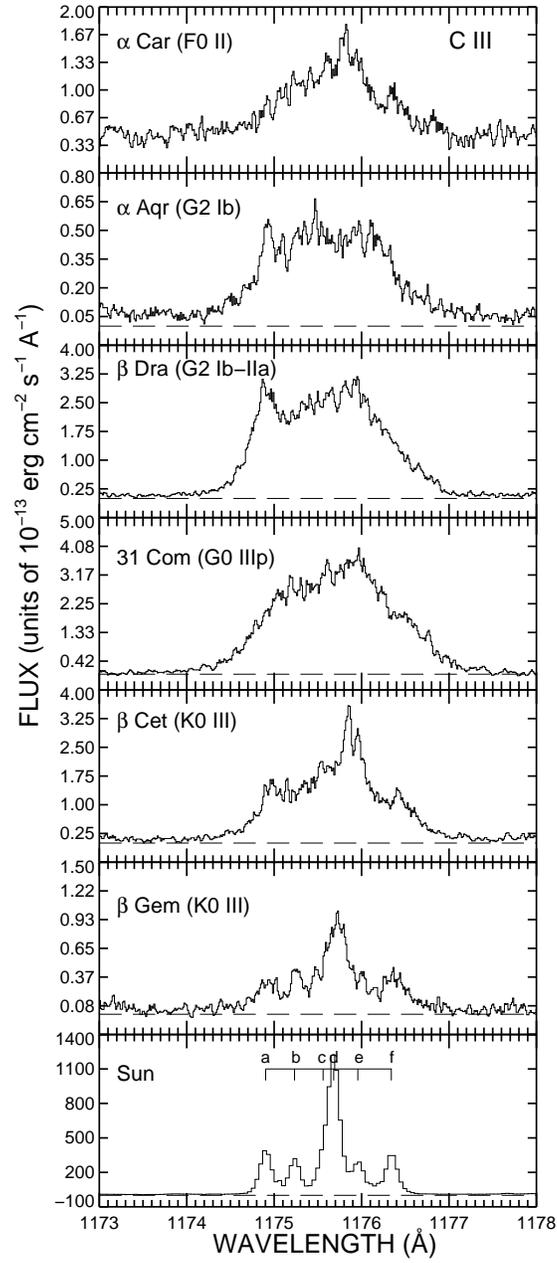}
\caption{The C III \gla  1176 multiplet in the target stars as
measured in the LiF2A channel. A solar sunspot spectrum 
(Curdt \etal\ 2001) is shown in the lowest panel.  The 6 components
of the multiplet are marked: {\it a .... f}.\label{c3.1176}}

\end{figure}

The \fuse\ region contains two strong transitions from
C~III, (\gla977 and \gla1176), whose ratio is principally sensitive
to electron density in optically thin plasmas 
over the range 10$^8$--10$^{11}$ cm$^{-3}$
(cf. Dupree \etal\ 1976).  These transitions have been widely
utilized in solar studies, and more recently in dwarf stars using
\fuse\ spectra (see
Redfield \etal\ 2002).  However if one or both of the lines are optically
thick, a simple ratio diagnostic can not be used.  Signs of optical
depth in the stellar \gla977 line were first noted in several 
targets from ORFEUS spectra indicated by anomalous widths and
relative fluxes (Dupree \& Brickhouse 1998).
Spectra from HST and \fuse\ illustrate optical depth effects as well 
(DelZanna \etal\ 2002, Redfield \etal\ 2002).
\fuse\ spectra of luminous stars reveal not only asymmetries in the
\gla977 line (see Sec. 11.2 and Fig.~\ref{c3.977}), 
but signs of anomalous ratios among components of the \gla1176
multiplet too (see Fig~\ref{c3.1176}).  As compared to the 
profile in the quiet Sun, this multiplet is compromised by  optical depth effects.
The profiles of \gla1176 in the stars  are generally not dominated
by the central component (\gla 1175.709, $2s2p\ {^3}P_2 - 2p^{2}\
{^3}P_2$) 
of the six transitions forming the multiplet as they are in 
the Sun.  The central transition shows the greatest effect
of optical depth where, near the solar limb, it becomes weaker
by as much as a factor of 2 (Doyle \& McWhirter 1980).
Only in the \fuse\ spectrum of \gb~Gem does the \gla1176 profile
appear to be optically thin, however there are signs  from profile
fitting and bisections that
the \gla977 transition in this star is not optically thin.
We conclude that {\it the \gla1176/\gla977 multiplet ratio can not be
applied to infer electron density in the chromospheres of these
luminous stars}. 

The relative strengths of certain members of the \gla1176 multiplet can
indicate electron density if they can be separated.  The 2--2 transition
(\gla 1175.709) is the strongest, and in the optical thin
case, its ratio (or the ratio of the
blend of \gla1175.709 [line d] and \gla1175.587 [line c]) 
relative to \gla 1175.983 (line e) is
sensitive to density over the range 10$^8$--10$^10$ cm$^{-3}$.
Excluding the strongest line from the multiplet, the ratio of 
\gla1175.983 (line e) to \gla1175.260 (line b), \gla1175.587 (line c) 
or \gla1176.367 (line f), could be used if sufficient signal
is available; they are also sensitive to density between
10$^8$ and 10$^{10}$ cm$^{-3}$. Only \gb\ ~Gem  appears to 
have an optically thin multiplet in which these ratios can be used.  
Using a profile of \gla1176 from the combined LiF2A and LiF1b segments 
for \gb\ Gem, we fit the six components of the multiplet simultaneously 
with Gaussians by adopting the laboratory wavelength separations and
holding all lines in the multiplet to the same full width at
half maximum.  The line ratios, $e/(c+d)$, $e/c$, $e/b$, and $e/f$ set a lower 
limit on  the electron density of 10$^9$ cm$^{-3}$ for a temperature  
T=80,000K using rates from CHIANTI (Young \etal\ 2003).  Spectra with longer
exposure times are needed to constrain a high density limit.

\section{Profile-Fitting}

Because some emission lines from cool stars do not appear
gaussian in shape, a practice (Wood \etal\ 1997) has developed to 
invoke multiple gaussian components to characterize
the line profiles.  The C III (\gla977) and O VI (\gla1032) lines
are the strongest stellar emission lines in these spectra and
most amenable to multiple component fits.  Our line-fitting procedure
is applied directly to the spectrum of photon counts because
this technique enables proper assessment of errors.

The reason for a preference for photon-fitting 
derives from the characteristics of the
spectrum and the \fuse\ detectors. The background level 
of  \fuse\ spectra is extremely low and,
coupled with the intrinsically low continuum levels of cool stars, results
in the spectra containing typically 0 to 2 counts per bin outside of
emission lines and in extended line wings.  Measurement errors 
for such low count levels are not
distributed according to Gaussian statistics, and so the minimization of
$\chi^2$ to derive emission line parameters is not appropriate for such
spectra (e.g., Nousek \& Shue 1989).

The method employed here is to minimize the $C$-statistic (Cash 1979)
which treats the statistics for small counts per bin data correctly. $C$
is defined as

\begin{equation}
C=2\sum_{i=1}^N ( f(x_i;a) - n_i \ln f(x_i;a) )
\end{equation}

\noindent
where $N$ is the number of data-points, $f(x_i;a)$ is the function fitted
to the data (dependent on parameters $a$), and $n_i$ is the number of
counts in bin $i$. In the present case the emission lines are treated as a
superposition of one or more Gaussians and a linear background.  Both
background and lines are fit simultaneously.
Minimization of $C$ is performed using Powell minimization through a
routine available in IDL (powell.pro).
Results of the fits for C III \gla977 and O VI \gla1032, are discussed
below.

\subsection{C III, \gla977}

The C~III  \gla977 line is affected
by interstellar absorption or central reversals in 
several stars, and for these stars we omit
from the fit those points affected by the absorption.
The C~III \gla977 transition is
fit by a single Gaussian profile.  
All of these single gaussians are shifted
to longer wavelengths which, if representing coherent
mass motions in the atmospheres would suggest the presence of infalling
material.  If symmetric emission is simply shifted 
to longer wavelengths,  the infalling emission region must  arise from the
whole atmosphere behaving coherently.  Of course  stars
can have complex surface structures with an uneven distribution
of regions of activity that produce departures from symmetry in
the line profiles.  In solar magnetic structures,  
a restricted  emission region produces  profiles 
characterized by red shifts in transition region lines  
(Doschek \etal\ 1976; 
Teriaca \etal\ 1999; Peter \& Judge 1999),\footnote{The SUMER
spectrograph on the ESA/NASA Solar and Heliospheric Observatory (SOHO) 
mission, used for most of these
measurements, has lower spectral resolution than \fuse,  -- a
2-pixel element covers $\sim$0.09\AA\ or R$\sim$12,000 so that
detailed shapes of the narrow solar emission line profiles can 
not be measured.}  
although anomalous center-to-limb behavior suggests other
mechanisms are present (Achour \etal\ 1995 and references therein).  
Dwarf stars also show redshifted emission in transition region
lines (Wood \etal\ 1997; Redfield \etal\ 2002).
In the luminous giant and supergiant 
stars observed with \fuse, it appears more likely to assume as
a working hypothesis  that there is 
opacity in the C~III \gla977
line resulting from outflowing material and 
causing the appearance of a red shift.
Radiative transfer effects (Hummer \& Rybicki 1968) can cause
the appearance of a red shifted profile resulting from increased
opacity on the blue side of the line.  Semi-empirical models
of luminous stars have demonstrated such asymmetries in chromospheric line
profiles (cf Lobel \& Dupree 2001).

The presence of such opacity can be
investigated by fitting a Gaussian profile {\it only} to the
long wavelength wing of the line, eliminating from the fit, both 
the peak emission and the short-wavelength side of the 
line profile from the peak to $\sim -$200
\kms.  These wing fits are also shown in Fig~\ref{c3.977} and 
parameters listed in Table~\ref{tbl.c3fits}.  The one-sided Gaussian fits
predict a line center that is less than 5 \kms\ from the predicted
photospheric velocity of all stars.
The resultant fits are consistent with the idea 
that the observed profiles are asymmetric
with the short wavelength side of the profile subject to 
absorption.  

\begin{figure}
\begin{center}
\includegraphics[angle=0.,scale=.75]{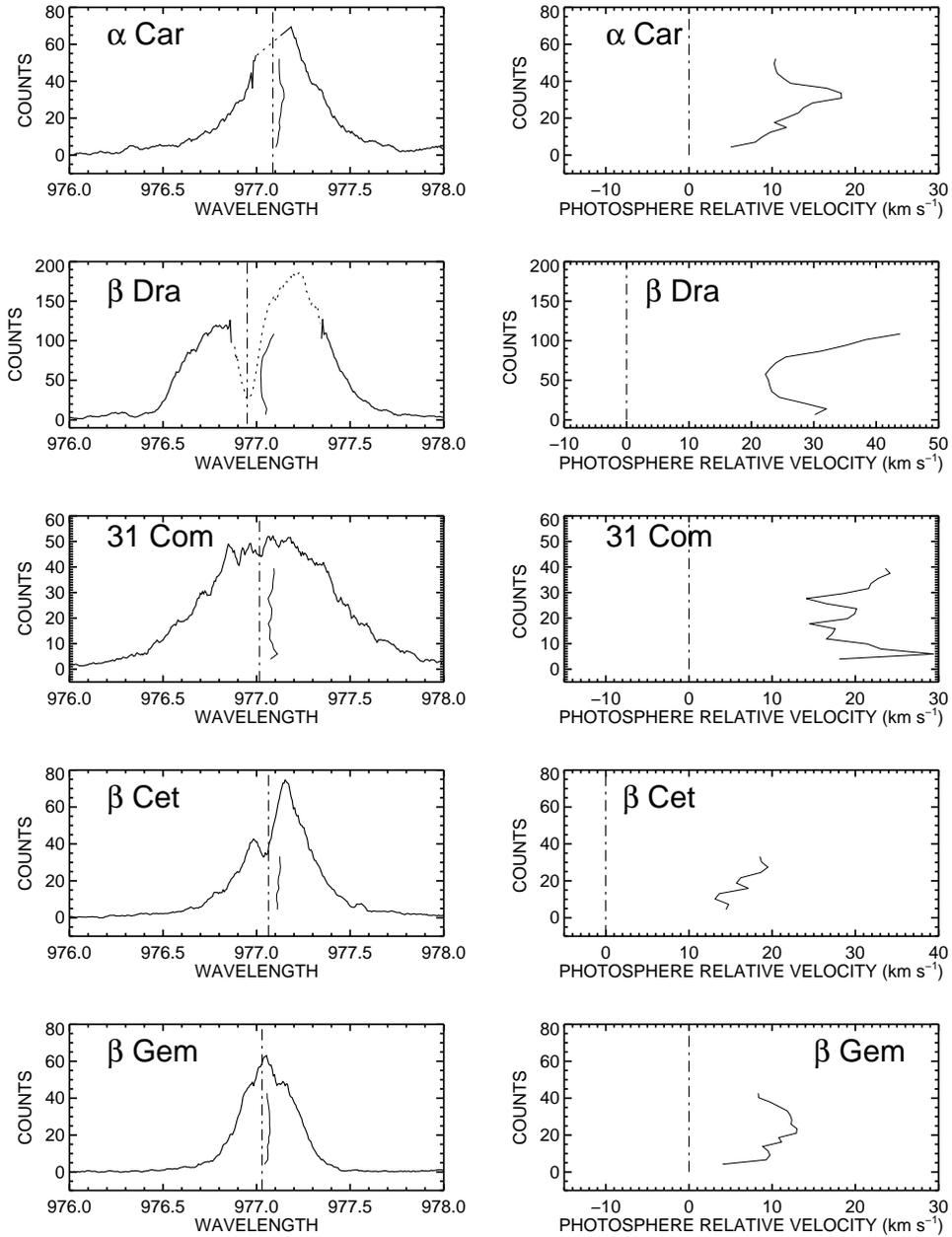}
\caption{Bisector of the C III \gla 977.920 emission in five targets.
The stellar profiles and bisectors are shown in the left 
panels, and the position of the bisectors are given on a velocity
scale in the right panel. 
Note that the bisectors do not extend below 5 counts so as not to be
affected by possible weak airglow emission.  Dashed line in \gal~Car 
and \gb\ Dra profiles indicates the region omitted from the bisector
process.  H$_2$ absorption on the short wavelength side of \gb~Dra
compromises the bisector below $\sim$50 counts.  Errors in
centroiding are less than 2 \kms. Alpha Aqr is not included because
the line is obviously absorbed on the short wavelength side (see
Fig. 9 and Fig. 20).\label{c3.bisect}}
\end{center}
\end{figure}

For several targets,  it is
useful to further characterize these line profiles using
a bisection technique.
The bisector of a symmetric emission line should 
remain at constant wavelength (or velocity) for all parts
of the profile.  To determine the bisectors, the profiles 
were smoothed, and cut into 25 segments, each
of flux strength 1/25 of the profile peak. The
weakest part of the profile ($\le$ 5 counts) was not included.  
The centroid of
each segment was determined omitting regions crossing the
interstellar absorption feature.  
The resulting bisectors of the C III \gla977 emission in  5 targets
(Fig.~\ref{c3.bisect})
show that these profiles are not symmetric.
Certain systematics are apparent from Fig.~\ref{c3.bisect}.  Towards
the base of the lines the bisector shifts toward negative  velocities
which could arise from geometric blocking by the
stellar disk at high positive velocities 
and/or decreased wind opacity at high
negative velocities causing line emission to appear.
In 31 Com, the core of the \gla977 line itself appears
asymmetric with enhancement at positive velocities, 
much like  the O VI \gla1032 profile in \gb\ Dra (see
following text).  Both profile
fitting and bisectors suggest that opacity is present in the
C III \gla977 profile in all targets except \gal\ Tau where the
signal is weak.

\subsection{O VI, \gla1032}
\begin{figure}
\plotone{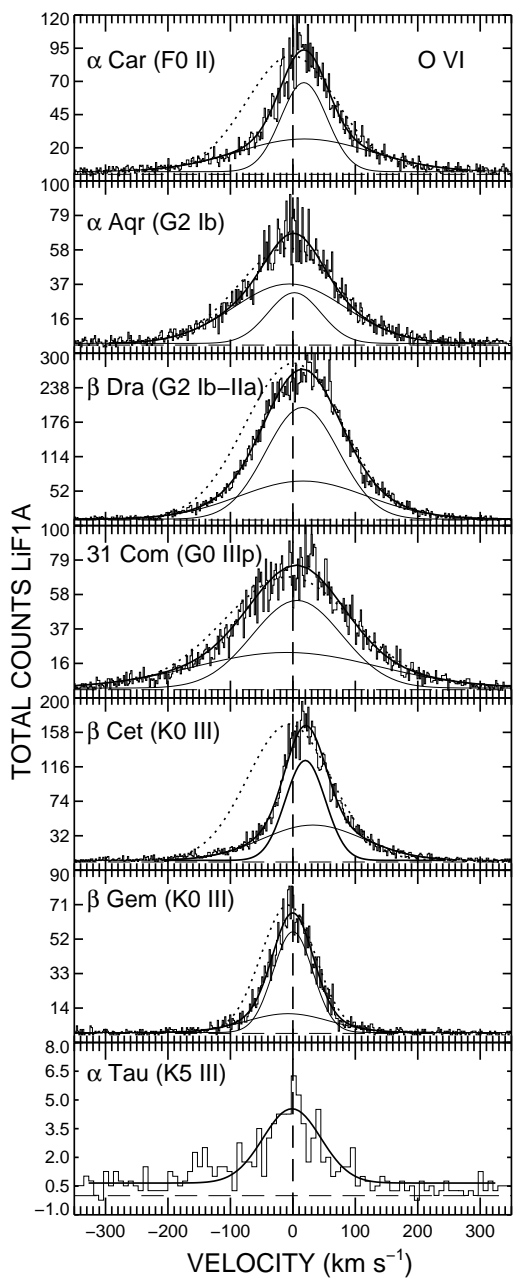}
\caption{The O VI \gla 1032 transition in the target stars. Except
for \gal\ Tau, multiple
Gaussians have been fit to the line profiles (indicated by the
thin solid lines) and the sum is marked by a thick solid line.   
Single gaussians have also  been fit to the
positive velocity side of the profiles (dotted line) to display
the intrinsic line asymmetries. 
The position of the stellar photosphere is marked by the broken line
at 0. \label{o6.1032}}

\end{figure}

The O VI \gla 1032.926 (Kaufman \& Martin 1989)
profile has extra emission in the wings of most targets, and as 
expected, 2 Gaussian  curves appear to 
produce a better fit to the profile than a single curve 
(see Fig.~\ref{o6.1032} and Table~\ref{tblprofiles}).
Centroids
of the narrow and wide Gaussian appear coincident in
\gal\ Car and \gb\ Dra, but they are separated in the
remaining 4 targets where it can be measured.  Three
of these stars (\gb\ Gem, 31 Com, and \gal\ Aqr) show a narrow
component that is shifted to longer wavelengths than the
wide component.  \gb\ Cet has the opposite shift: the
narrow component is shifted to shorter wavelengths.  
The ratio of widths of wide:narrow vary between a factor of 1.7 to 2.9.
The flux in the narrow component is generally larger by a factor of
1.1 to 2.6 than 
the flux in the broad component except for \gal\ Car (0.94) 
and \gal\ Aqr (0.40).

\begin{figure}
\includegraphics[angle=0.,scale=0.8]{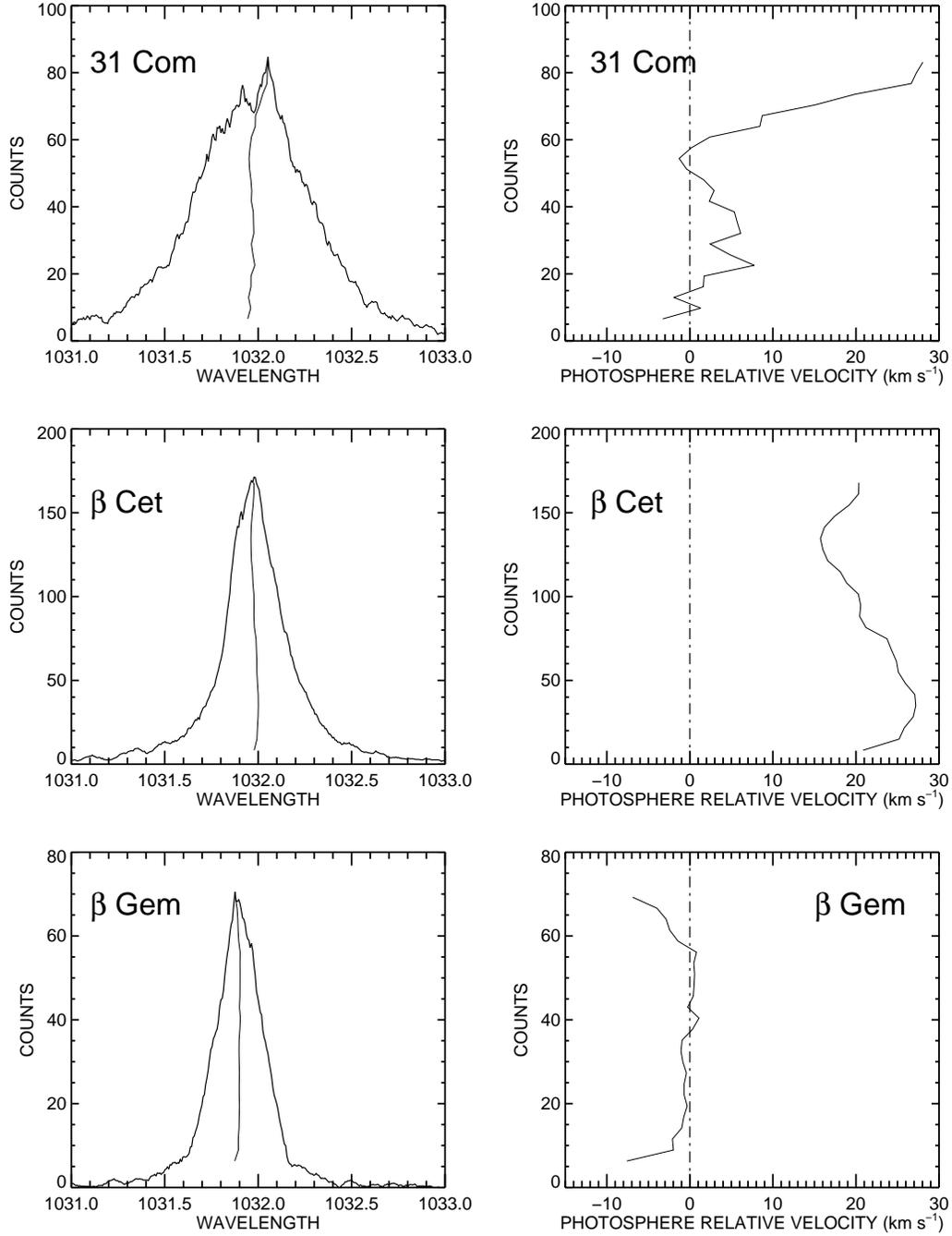}
\caption{Bisectors of the O VI (\gla1032) emission in the giant
stars. With the \fuse\ spectral resolution of 13--17 \kms, errors
in centroiding are less than 2 \kms.\label{bisectorsb}}

\end{figure}

\begin{figure}
\includegraphics[angle=0.,scale=0.8]{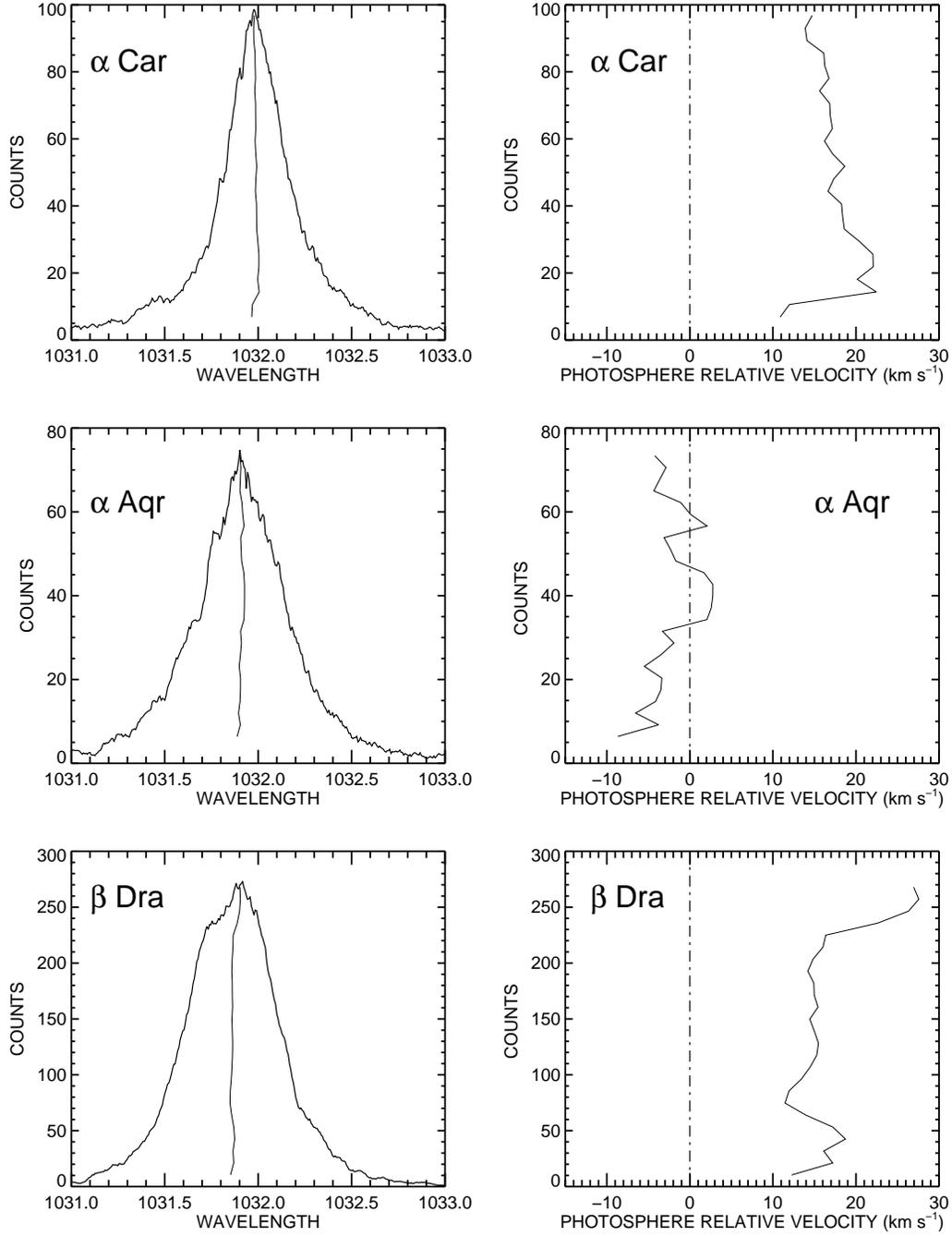}
\caption{Bisectors of the O VI (\gla1032) emission in the supergiant
stars.\label{bisectorsa}}

\end{figure}

The physical interpretation 
of a 2 Gaussian fit to \gla1032 is not obvious (see Sec. 11.2).
Fitting the long wavelength side of the profile with a single
Gaussian is also shown in Fig.~\ref{o6.1032} and Table~\ref{tblo6.red} 
with the exception of
\gal\ Tau. Evidence for absorption
on the short wavelength side appears in all cases.
A line bisecting the smoothed O~VI, \gla1032 profile
(Fig.~\ref{bisectorsb} and Fig.~\ref{bisectorsa}) demonstrates asymmetries
in all stars.  The line cores also merit notice.  Line center,
where the optical depth is largest, might be expected
to show the first signs of opacity.  Beta Dra and 
31 Com show the greatest velocity variation of the bisector 
at the peak of the profile.  The cores are asymmetric with extra
emission on the long wavelength side.
We suggest this is another sign of opacity and  
outflow. 

Information is provided by a direct comparison of
members of the O VI multiplet.  Since  
oscillator strengths of \gla1032/\gla1037 are in the ratio 2:1, the
\gla1032 line has a larger optical depth than \gla1037. 
Overlaying the actual \gla1032 profile (divided by 2)
on the \gla1037 spectrum, we find the short wavelength wing
of \gla1032 lies {\it below} the corresponding short wavelength side
of the \gla1037 profile (except for Beta Dra which has 
substantial H$_2$ absorption.)  This lends additional support to the presence
of opacity in O VI.

\section{H$_2$ Absorption}

\begin{figure}
\begin{center}
\includegraphics[angle=0.,scale=0.8]{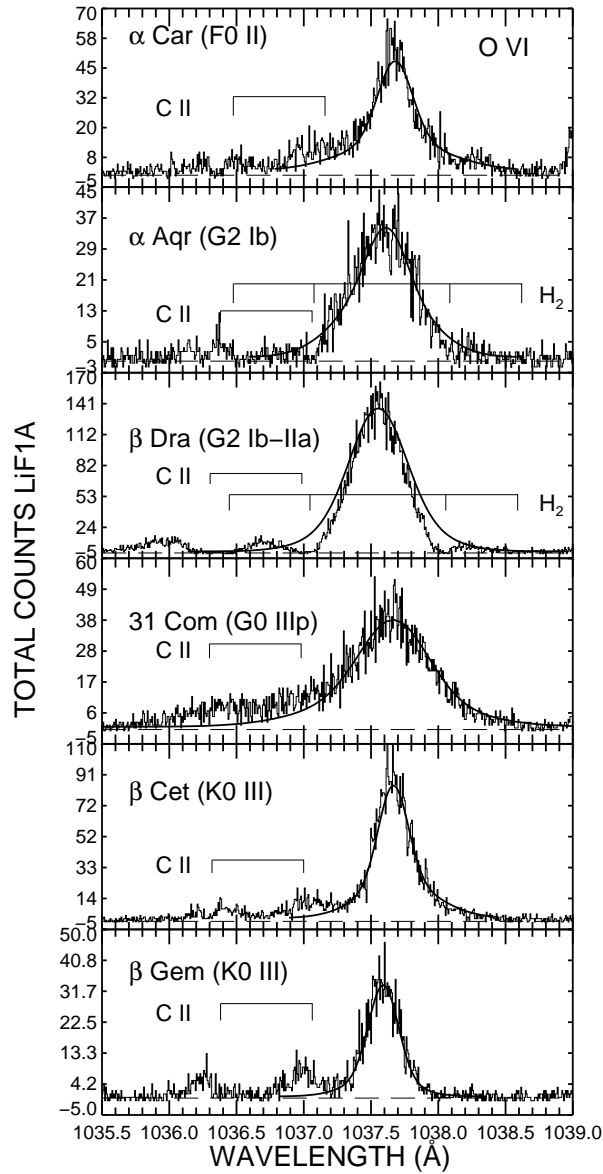}
\caption{The \gla1037 O~VI profile in target stars.  A 2-Gaussian fit to the
corresponding O~VI \gla1032 line profile is overlaid, 
scaled by 0.5.  Clearly \gb\ Dra and \gal\ Aqr give evidence for
H$_2$ absorption indicated by absorption near the base of the
profile.  The C II emission doublet is present in many stars, but can
be affected by H$_2$ absorption.\label{h2abs}}

\end{center}
\end{figure}

The weaker component of the O VI multiplet,  \gla1037.617, is shown in 
Fig.~\ref{h2abs}.  Nearby is C II emission.  The scaling of the
2 Gaussian fits to the 1032 line, reduced by a factor of 2
representing
the optically thin ratio, gives a reasonable fit to the \gla1037
emission although usually differs in detail.  
This procedure illustrates the presence of 
absorption by interstellar H$_2$ near the 
\gla1037 line in the spectra of  \gb\ Dra and \gal\ Aqr (Fig.~\ref{h2abs}).  
The  corresponding  H$_2$ 
transitions are shown in the appropriate
panels of Fig.~\ref{h2abs}.  The position of  H$_2$ absorption
was computed from the files made available by S. R. McCandliss
as $H2ools$ on the
\fuse\ website ({\it http://www.pha.jhu.edu/$\sim$stephan/h2ools2.html}), 
for a column density of 10$^{18}$ cm$^2$, T=100K, 
and b = 5 \kms\ (McCandliss 2003).
It is not surprising that these 2  stars in 
our sample, being among the most distant, 
show evidence of $H_2$ absorption.  These two stars are
near the plane of our Galaxy and located in the single sector where $H_2$ absorption
has been detected by \fuse\ in the spectra of white dwarfs
between 100 and  200 pc distant (Lehner \etal\ 2003).  
Lehner \etal\ suggest that
the $H_2$ in the local interstellar medium may occur as one large
diffuse cloud, possibly an extended thin sheet.

\section{Discussion}

\fuse\ spectra show that warm atmospheres, with temperatures
up to and including 300,000K (the temperature of formation of
O VI) are present in all stars, except the M supergiant, \gal\ Ori
(see Fig.~\ref{targets}).  Alpha Tau is particularly 
interesting because X-rays have not been detected from this
star (H\"unsch \etal\ 1996) yet  there is 
clearly high temperature (3 $\times$10$^5$K) 
plasma in the atmosphere.  By analogy with solar coronal
holes, the coronal temperature could be less where the
high speed wind originates and a high temperature stellar corona
might not be present where there is a strong wind.  The supergiant
\gal\ Ori is an obvious extreme example.   In the Sun, the underlying energy flux
is comparable between closed and open magnetic regions (Withbroe 
\& Noyes 1977), but
in open regions, the energy goes into driving the wind, and not
into heating the atmosphere. 

\subsection{Systematic Flux Variations}

\begin{figure}
\includegraphics[angle=0,scale=1.]{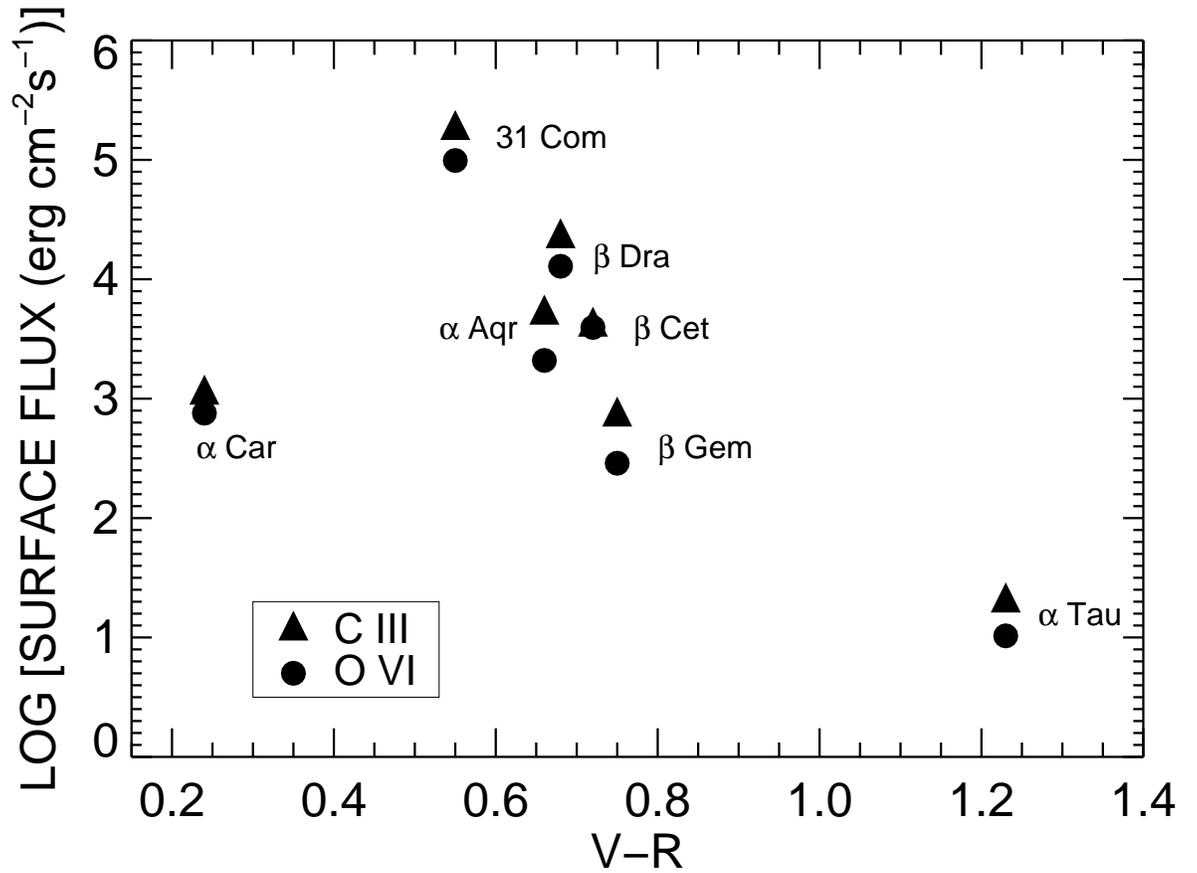}
\caption{Emission line flux in C III \gla 977 and O VI \gla 1032 
at the stellar surface as a function of (V$-$R).  Uncertainties in
the absolute flux are $\sim$10\% which are about  the size of the plotted
symbols. \label{surface.flux} }

\end{figure}

\begin{figure}
\includegraphics[angle=0,scale=1.]{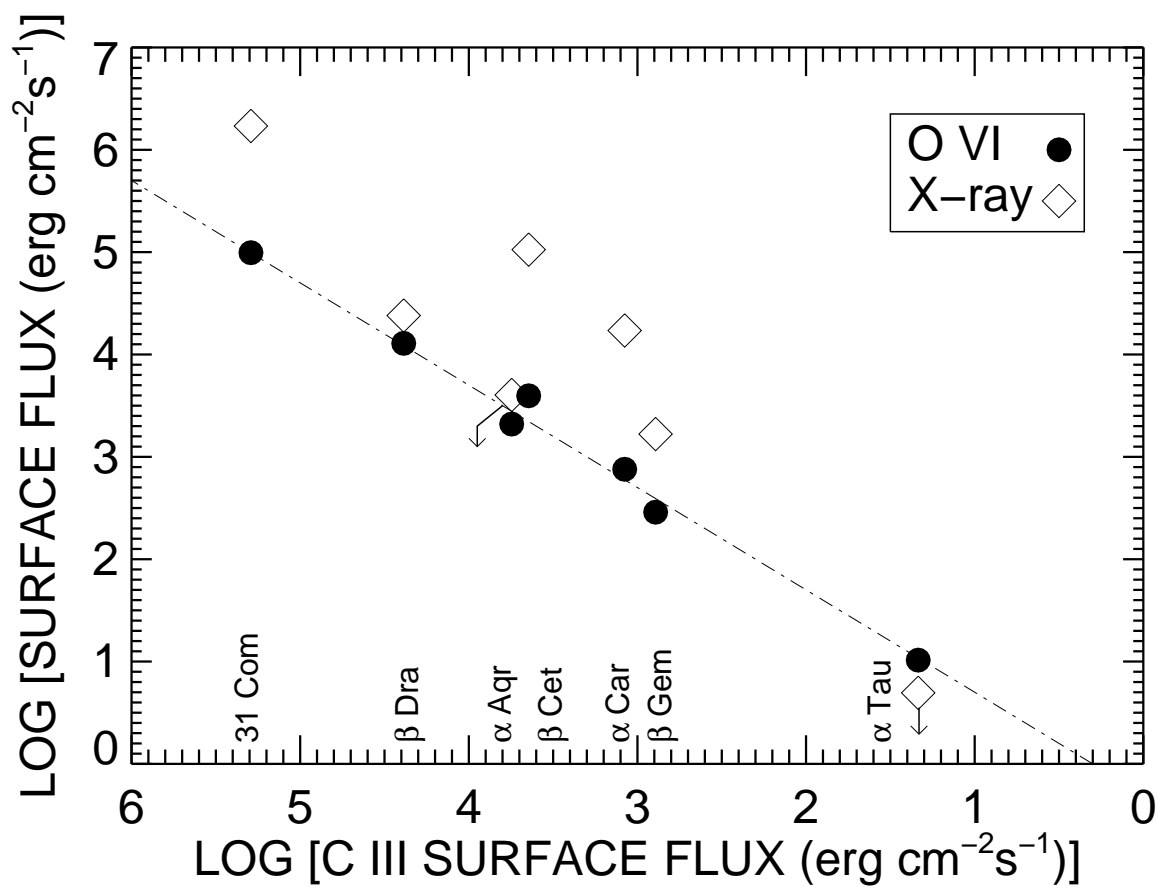}
\caption{The tight correlation between C III (\gla 977) and
O VI (\gla 1032) stellar surface flux ({\it filled circles}) 
contrasts with the
scattered relation between X-ray and C III flux ({\it open diamonds}).
X-ray measures taken from the ROSAT PSPC (0.1--2.4 keV) as 
reported by Ayres \etal\ (1995). X-Ray values from H\"unsch \etal\ (1996)
for \gal\ Tau and H\"unsch \etal\ (1998) for
\gal\ Car are shown.  X-ray fluxes for \gal~Aqr and \gal~Tau are
upper limits. Errors on the measured fluxes of C III, O VI, and the
X-ray flux are $\sim$10\% or less.\label{correlation}}

\end{figure}

\begin{figure}
\includegraphics[angle=0, scale=1.]{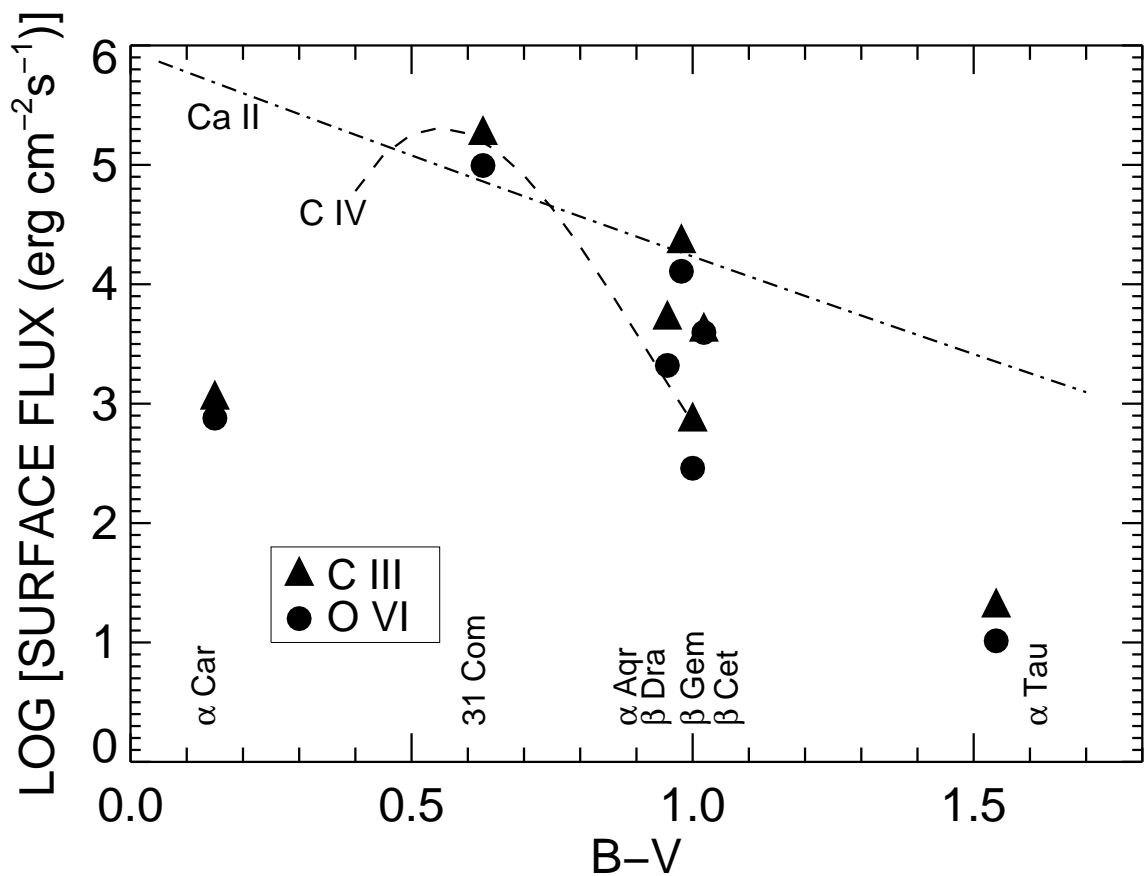}
\caption{Surface fluxes in C III(\gla 977) and O VI(\gla 1032)
as a function of (B$-$V) for comparison to behavior of other
emissions in giant stars. The net flux in Ca II (H $+$ K) 
in open cluster giants (Dupree \etal\ 1999) 
and the C IV (\gla 1550) flux (Ayres \etal\ 1995) are denoted
by {\it dot-dash} and {\it broken} lines respectively.  Errors in
measured fluxes are $\sim$10\% -- comparable to the size of the symbols.\label{flux.compare}}

\end{figure}

The surface fluxes of the C III and O VI lines for each
star were calculated using the Barnes-Evans relationship 
(Barnes, Evans, \& Moffett 1978)
between $V-R$ color and surface brightness.  The surface
fluxes decrease systematically towards lower effective
temperatures (Fig.~\ref{surface.flux}).  The C III and
O VI fluxes are tightly correlated (see Fig.~\ref{correlation})
and, by contrast the X-ray flux exhibits more variation suggesting
different heating mechanisms.  The decay of the transition
region lines appears similar to that of C IV in many giant stars,
and distinct from the chromospheric Ca II behavior
(Fig.~\ref{flux.compare}).
The rapid decay of the transition region emissions (C III,
C IV, and O VI) with temperature
is reminiscent of models of magnetic dynamo behavior (Rutten \&
Pylyser 1988; Dupree \etal\ 1999) in contrast to acoustic
models (Buchholz \etal\ 1998).

The enhancement of surface emission of the transition lines
in 31 Com, a rapidly rotating giant, over that in other giants
is consistent with formation by a  magnetic dynamo process
which becomes vigorous in fast rotators.

\subsection{Line Profiles}

Displaced Gaussian profiles have been observed in transition
region emission lines in the Sun (Doschek \etal\ 1976; Peter \& Judge 1999; 
Teriaca \etal\ 1999), and a wide variety of
cool stars (Wood \etal\ 1997; Redfield \etal\ 2003, and this paper).
In the Sun, when observing a restricted atmospheric region 
($\sim$ 1--2 arcsec in size), gaussians shifted
both to shorter and longer wavelengths are found.  The source
of the red-shifted emission has been attributed to many causes
(unidirectional mass flows along magnetic loop structures, microflaring,
heating effects etc.), but no definitive identification of the causes
of the redshifts has emerged (Peter 2004).

\begin{figure}
\includegraphics[angle=0,scale=1.]{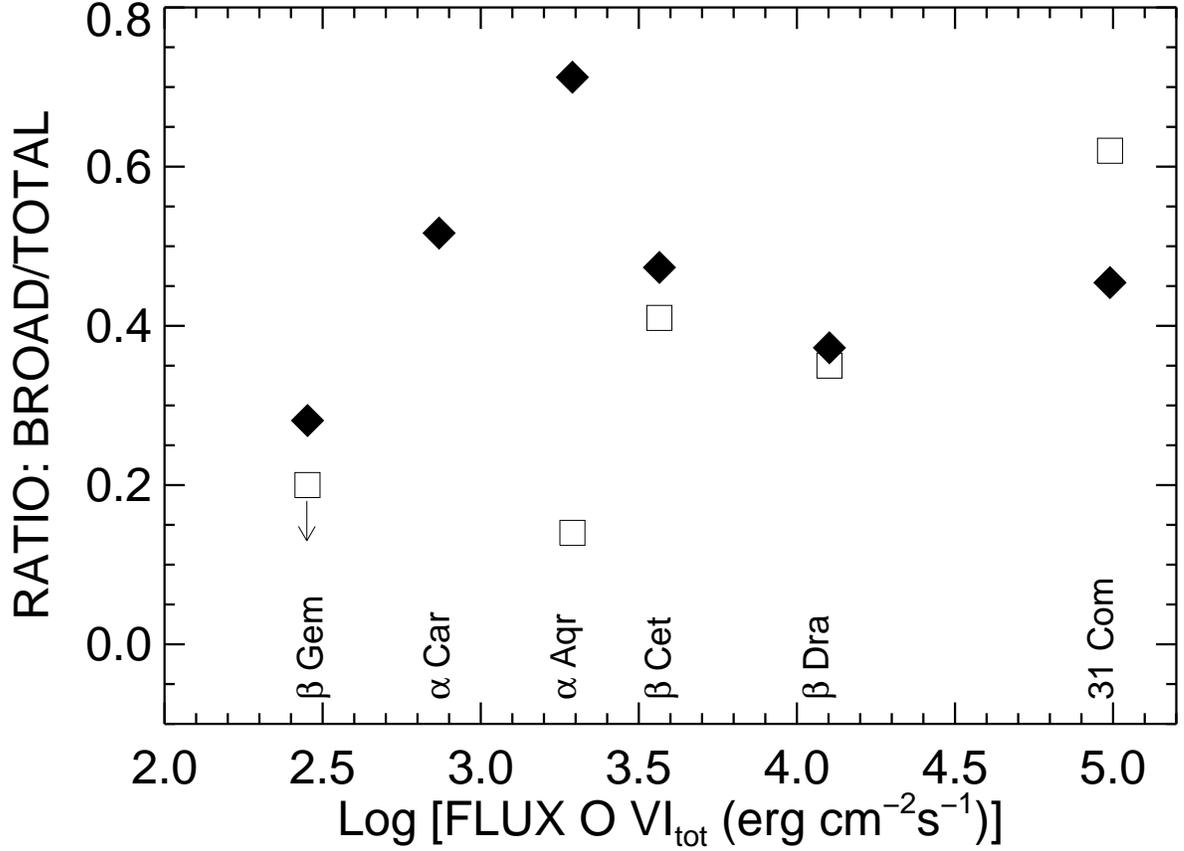}
\caption{Ratio of the flux in the broad component 
of O VI   to the total surface 
flux in the O VI \gla1032 line ({\it filled diamonds}) 
as measured from Gaussian fits to the profiles.  Results for some of 
the same stars from HST spectra of  C IV \gla1550 are shown 
({\it open squares}, from Wood \etal\ 1997). The C IV ratio for \gal\ Aqr
was measured from an HST/GHRS archival spectrum.  Formal errors in
these ratios are on the order of $\pm$10\% or less.  Formal errors in the
\fuse\ flux measurements are comparable to that value.\label{o6.broadflux}} 

\end{figure}

The origin of the broad component in stellar line profiles of Si IV
and C IV has been ascribed to microflare heating of
the transition region (Wood \etal\ 1997)  
much as ``explosive events'' occur on the
Sun.  Additional evidence for this conjecture 
has been offered by the increase in the relative
contribution of the  broad component to the
total flux accompanied by an increase in
the C IV line flux and X-ray flux.  This correlation suggested 
that enhanced heating 
contributes to the broad profile (Wood \etal\ 1997).  The C III profiles
studied here do not require two Gaussians (however their
distinctive asymmetries are discussed below).   Moreover, the
2 components of O VI 
do not behave similarly to C IV in all  stars 
(see Fig.~\ref{o6.broadflux}) and
show no systematic dependence on activity level.
Thus, for luminous stars the behavior of a broad component
does not appear similar to that identified in dwarf stars
(Redfield \etal\ 2002).  If such a relationship exists, it appears
to be confined to the Si IV and C IV lines, as found by Wood \etal\
(1997) and does not extend to lower or higher temperatures.

Detailed studies of the C IV profiles in the Sun show that
2-component fits are required only in the network regions,
and not in the internetwork areas where 1 Gaussian suffices to match
the profile. Moreover
emission from explosive events or transient brightenings
is not related to the  broad wings in the Sun (Peter \& Brkovi\'c
2003). A remaining
possibility  to interpret 2-Gaussian fits appeals 
to the geometry of the transition
region. Peter (2001) suggests that the broad wings 
on the solar disk originate
from Alfv\`en wave-heated coronal funnels that accelerate
the solar wind.  Such broad wings are very apparent when viewing
sections of the solar corona  along a line of
sight that traverses a coronal hole where a fast acceleration
occurs.  Outflow velocities
$>$200 \kms\ significantly broaden the line profile (Miralles \etal\ 2001).

The luminous stars in this survey have extended atmospheres due in
part to their low effective gravity,
and it appears possible that much of the line broadening may be attributed
to extension and expansion. This was first suggested by the
observation of broad emission lines in hybrid stars as measured
with IUE (Hartmann \etal\ 1981).  

Analysis of the line
profiles for dynamical signatures offers an explanation. 
The shapes of emission line profiles can give clues to the
atmospheric dynamics through the presence of
line asymmetries. As Hummer \& Rybicki (1968)
first noted (and more recently Lobel \& Dupree 2001),  
a differential expansion (or contraction) can
cause red (or blue) asymmetries of the line profile.  In complex
multicomponent atmospheric modeling, when spatially averaging the
contributions
from many effectively-optically thin components, each with
potentially different velocity structures, the resulting
line profile will be an appropriately weighted sum of the
contribution
function of each component.  

The opacity at line center, for a thermally 
broadened line is proportional to 
$A_{Z}\times \lambda \times f \times (M/T)^{1/2}\times N_e$
where, $A_Z$ is the elemental abundance with respect to
hydrogen; \gla, the wavelength of the line; $f$, the line oscillator
strength; $M$ the mass of the atom; $T$ the temperature of formation;
and $N_e$ is the column density of electrons over the line forming
region.  Although we do not yet have detailed models of the
atmospheres of these stars, the
atomic physics alone suggests that  of the two major emission 
lines in the \fuse\ region,
C~III (\gla977) and O~VI (\gla1032), the carbon line  
should have higher opacity (cf. Harper 2001 also). This 
amounts to a factor of 2.5 for values
of $A_Z\times \lambda \times f$ alone.  It is expected
that the electron column density will be higher for the
C~III line forming region than for that of O~VI since the emission
measure
distribution at C III temperatures exceeds that found at
the higher temperatures of O VI (cf. Sanz-Forcada \etal\ 2003).  
Thus we might
expect that the \gla 977 profile would be more sensitive
to dynamics.

Inspection of the C III line profiles (cf. Fig.~\ref{c3.977})
shows a broad line, usually crossed by interstellar absorption,
that is clearly asymmetric, displaying a lower flux at
negative velocities than at positive velocities.  This is obvious  
in the spectra of   \gal\ Car, \gal\ Aqr,  \gb\
Dra, and \gal\ Tau.  Similar (although less pronounced) asymmetries
are found in the C III profiles of \gb\ Cet and \gb\ Gem
as illustrated by the single Gaussian fits to the emission.
In both stars  
the line is asymmetric, suggestive of
excess opacity in the line at negative velocities. The \gla977
line from the fast rotating giant 31 Com possesses a FWHM $\sim$ 255 \kms.
The $vsin\ i$ of this giant (57 \kms, Strassmeier \etal\ 1994) is
about 
half the observed line width, so clearly
a line broadening mechanism in addition to
rotation is present, perhaps extension of the atmosphere.  
The Gaussian fit to the
long wavelength wing of the profile suggests that additional
opacity is present in 31 Com.

The O VI \gla 1032 profiles appear more symmetric than those
of C III \gla 977.  This might be expected because the opacity is
less than in the carbon resonance line.  It is not 
straightforward to predict the effects of opacity on the line 
profile.  A higher
optical depth is expected near the line center simply because
the absorption profile reaches a maximum; thus it is reasonable
to expect the core to show signs of opacity.

All of the O VI lines (except for $\alpha$ Tau which remains
indeterminate because the count level is low) show
an asymmetric profile.  Two characteristics are apparent, irrespective
of the absolute offset:
three stars (\gb\ Dra, \gb\ Cet, and 31 Com) show
a positive shift at the top of the profile (similar to that
found in C III \gla977) indicative of increased
opacity on the short wavelength side.  This is the signature of
absorption produced by outward moving material in these stars.

All bisectors exhibit a shift to negative velocities towards the base.  Two interpretations
for this behavior appear plausible.  Either the opacity in a
wind decreases as the expansion velocity approaches 100 \kms,
or the extended stellar atmosphere blocks the extreme
outward velocity of the atmosphere, creating a shift of the centroid
to shorter wavelengths, or both.

Line widths are informative as well.  Both \gal~Car and \gal~Aqr
show exceptionally narrow core profiles, when compared to
stars of similar luminosity, for instance \gb~Dra.  
And as discussed further in the following section, the O~VI width  in 
most objects is comparable to that of the  C IV line.

\subsection{Comparison with Ultraviolet Emission Lines}

\begin{figure}
\includegraphics[angle=0.,scale=0.8]{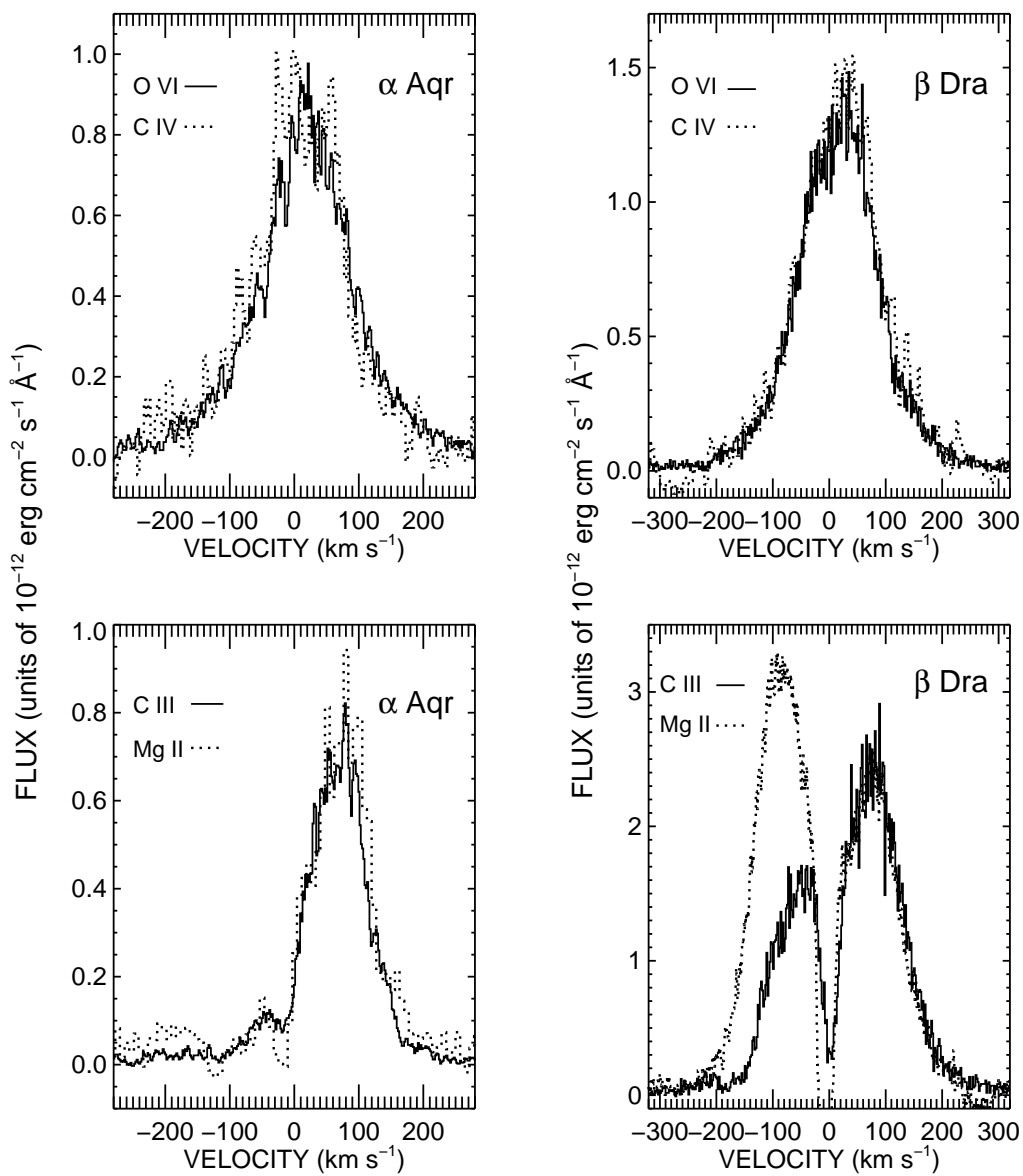}
\caption{Comparison of C III and O VI from \fuse\ with 
other emission lines in the
supergiants \gal\ Aqr and \gb\ Dra. For \gal\ Aqr, C IV (1548\AA) taken from 
HST/GHRS Dataset Z1FG010AM; Mg II (2795\AA) from IUE LWR01390. For \gb\ Dra,
C IV (1548\AA) taken from HST/GHRS Dataset Z2NW010CT; Mg II (2795\AA) from 
HST/GHRS Dataset Z0WZ0109T.\label{fuse.hst.1}}

\end{figure}

\begin{figure}
\includegraphics[angle=0.,scale=0.8]{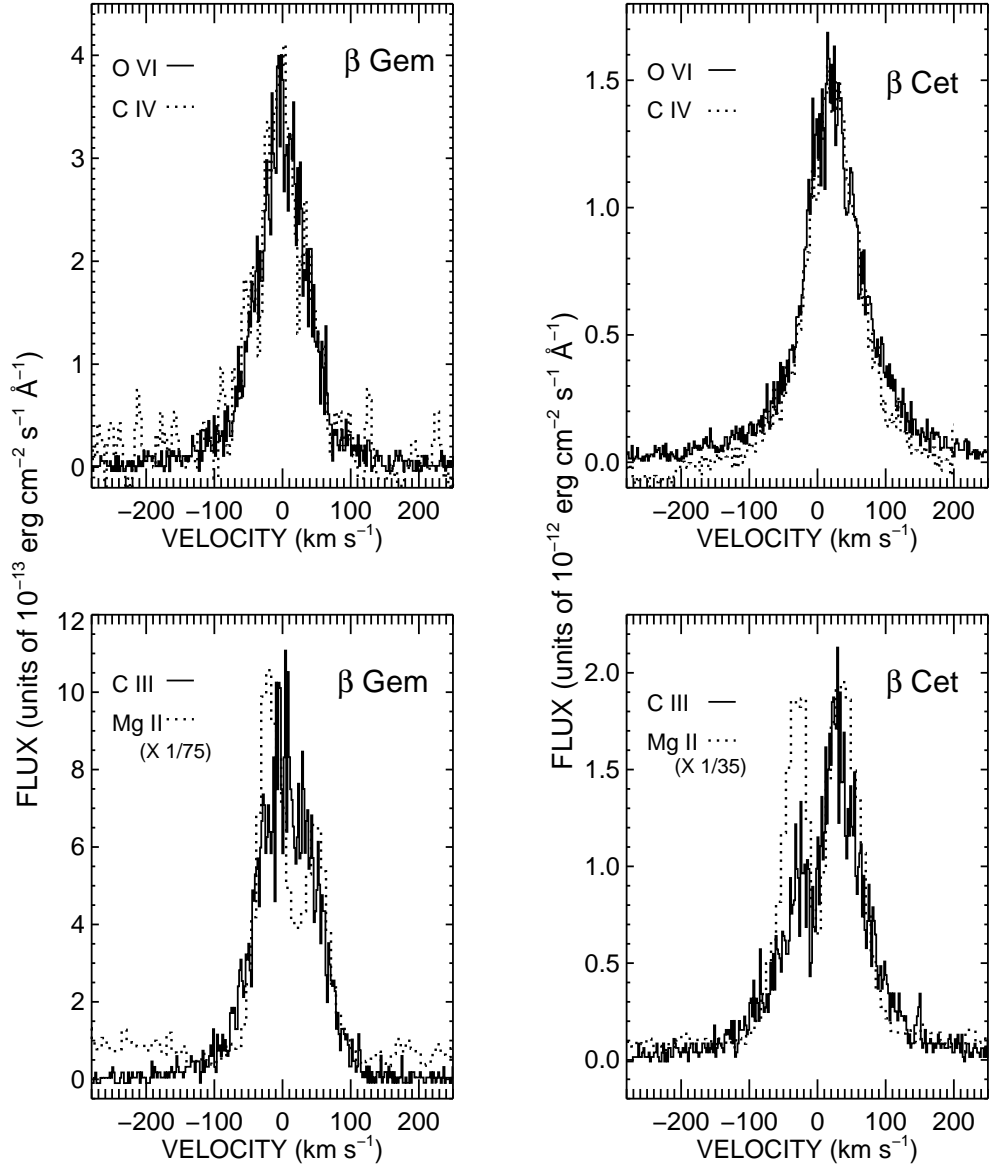}
\caption{Comparison of C III and O VI from \fuse\ with other 
emission lines in the giants \gb\ Cet and \gb\ Gem.  For \gb\ Cet,
C IV (1548\AA) taken from HST/STIS Dataset O5B701020; Mg II (2795\AA) from IUE LWP 08615.  For
\gb\ Gem, C IV (1548\AA) taken from HST/GHRS Dataset 
Z2UZ010FT; Mg II (2795\AA) from IUE LWP 27482. \label{fuse.hst.2} }

\end{figure}

\begin{figure}
\begin{center}
\includegraphics[angle=0.,scale=0.8]{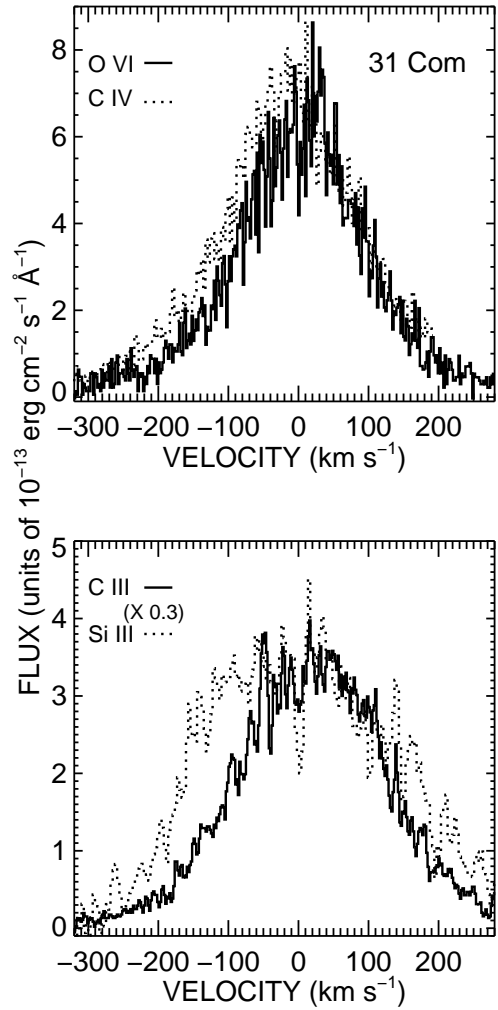}
\caption{Comparison of C III and O VI from \fuse\  
with other emission lines in 31 Com. The C IV (1548\AA) profile is taken
from HST/STIS Dataset O6AQ01010; Si III (1206\AA) from HST/STIS Dataset 
O6AQ01020.\label{fuse.hst.3}}
\end{center}
\end{figure}

\begin{figure}
\begin{center}
\includegraphics[angle=0.,scale=0.8]{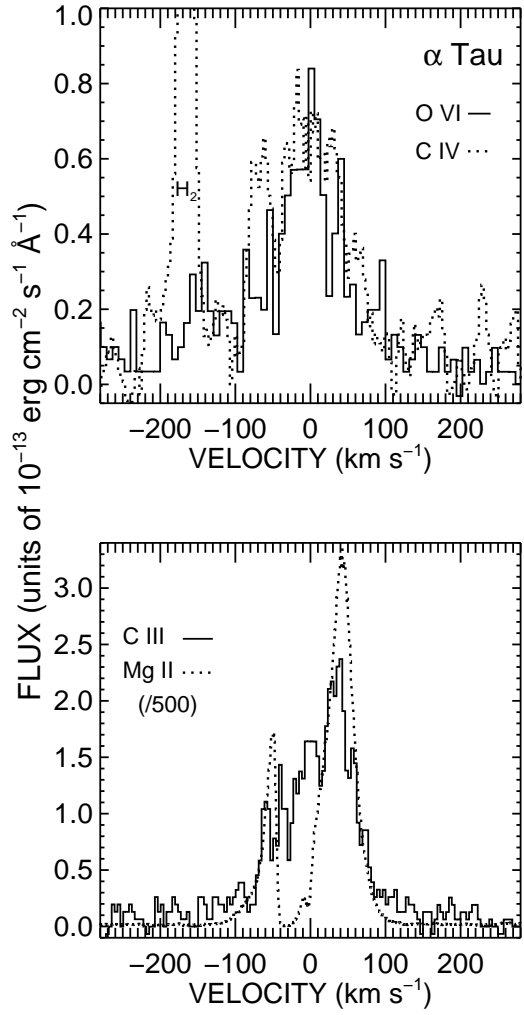}
\caption{Comparison of C III and O VI with other emission lines in
$\alpha$ Tau. The C IV profile is from HST/STIS Dataset  O6JE01020; the
Mg II profile from HST/GHRS Dataset Z3FX020BT.\label{fuse.hst.4}}
\end{center}
\end{figure}

The \fuse\ spectra of O VI sample the highest temperature transition 
region lines for which many line profiles are available; the C III profile
is the most sensitive to optical depth.  It is of interest
to trace the atmospheric dynamics by comparing emission
from C IV, Si III, and Mg II to the \fuse\ profiles. Profiles of the supergiants
\gal\ Aqr and \gb\ Dra are shown in Fig.~\ref{fuse.hst.1}. Whereas
\gal\ Aqr shows good agreement between the asymmetry of the
C III (\gla 977) line and the Mg II, indicating 
outflow and absorption at velocities up to $-$100~\kms or
more, \gb\ Dra spectra indicate that the outflow does
not occur at the cooler levels of Mg II, but at the
higher temperatures represented by C~III.  Variable opacity has
been noted on the short wavelength side of the C IV line
in \gb\ Dra (Wood \etal\ 1997) which provides evidence for
a wind at transition region temperatures.  It is thought
that \gb\ Dra may be in a pre-hybrid phase.   Alpha Aqr is
a well-known `hybrid' star where the wind is well developed
and detectable throughout the atmosphere; supersonic acceleration
has even been identified in the chromosphere (Dupree \etal\ 1992).  The C IV
and O VI line are similar in width in both supergiants and the
asymmetry
of the peak of the core emission persists in C III, C IV, and O VI 
in \gb~Dra.
Comparison of profiles for the giant stars are shown in
Fig.~\ref{fuse.hst.2}, Fig.~\ref{fuse.hst.3} and
Fig.~\ref{fuse.hst.4}.  
With the exception
of \gb~Gem, the
other giant stars, \gb~Cet, 31 Com, and \gal~Tau show opacity
in the C III line which is frequently matched
with a similar asymmetry in Mg II or Si III.  Beta Gem is
more like \gb~Dra with chromospheric infall indicated by the Mg II line shape.  
Since these spectra
come from diverse sources, including IUE, HST/GHRS and HST/STIS,
they have been scaled in flux and at times shifted  to
compare profile shapes.

It is puzzling that the O VI (\gla1032) line widths are generally comparable 
to the C IV (\gla1548) transitions and both are broadened in excess
of pure thermal broadening.   In the solar network, O VI is 
observed to be broader than the C IV  
line (Peter 2001), by a factor of 1.2 to 1.4; the non-thermal 
contributions are also higher in the Sun for
O~VI than C~IV.  
The observed O VI line width of the core 
exceeds the thermal width expected at 3$\times$10$^5$K by factors
of 3 or more.   Clearly atmospheric extension, turbulence, and
or opacity can affect these line profiles.

Could the character of the atmosphere change dramatically 
at the $\sim$200000K level from a relatively homogeneous outflow
(indicated by the broad asymmetric C III profiles) to an atmosphere covered
by magnetic loop structures signaled by the narrow redshifted C IV and
O VI lines?  We have no estimate of the densities in the regions forming
C~IV and O~VI so it is unclear whether small scales are
indicated (as they are for dense coronal material). Lower turbulent 
velocities and/or less geometrical broadening might be
plausible in such structures although they are not required for 
confinement because  thermal motion of material at the temperature of 3$\times$10$^5$K 
(18 \kms) is an order of magnitude less than the escape velocity from
giant stars (200 \kms).
As noted earlier, identifying the redshifted emission profiles of O VI with
physical atmospheric downflows requires synchronous motion among 
all putative loops covering the giant or supergiant stars, or
a judicious combination of many regions with individual 
dynamics that systematically produce a redshifted line.  
We can not firmly eliminate some distribution of emitting
regions over the stellar surface that produces a red-shifted asymmetric
emission line profile.  However the identical nature of the C IV and
O VI line profiles (measured at different times) suggests they are
not dominated by transient active regions or varying downflow emission
profiles.

Could wind opacity effects narrow the lines causing 
O~VI and C~IV to be less broad  than C III?  
Without a detailed model, it is
difficult to assess the relative opacities in the C~IV and O~VI 
transitions.  A comparison of the quantity, $f\lambda \times A_{el}$
using solar abundances suggests that C~IV opacity values exceed O~VI
by 20\%.  However, luminous stars are evolved, and it is well known
that the CN cycle depletes Carbon (enhancing Nitrogen).  
Although classical  studies suggest that
both carbon and oxygen are underabundant with
respect to solar values (Luck \& Lambert 1985), there is currently
controversy about the difficult-to-measure oxygen abundances
(Fulbright \& Johnson 2003).   The line shapes strongly suggest that 
opacity plays a role; clearly modeling is needed.

\subsection{Relationship to Coronal Lines}

Two or our targets show coronal line emission in the \fuse\ region,
and others are X-ray sources.  The \fuse\ spectra suggest that the
coronal lines occur near photospheric radial velocities, and so do not
participate in any outflow (also see Redfield \etal\ 2003).  
This is consistent with the fact
that plasma at coronal temperatures (here $\sim$6$\times 10^8 K$) must
be confined by magnetic fields in these stars.  Such confinement
was postulated when Fe XVIII and Fe XIX species  were first identified 
as a stable feature in the EUVE spectra of  the giant
stars of Capella (Dupree et al. 1993; Young \etal\ 2001).  Confinement is consistent with
the small sizes inferred from the high densities of the coronal  regions
at these temperatures (Sanz-Forcada \etal\ 2003).   These \fuse\ results
suggest
an inhomogeneous atmosphere in which small magnetic features at high
temperature are anchored in the presence of a warm expanding
atmosphere.

\section{Conclusions}

\begin{enumerate}
\item The presence of warm 3$\times$10$^5$ K plasma, indicated
by O VI emission appears ubiquitous and extends across the
HR diagram.  The K5 giant, \gal\ Tau is the coolest giant 
to exhibit O VI known to date.

\item The atmosphere of the M supergiant - \gal\ Ori
does not exhibit any C III or O VI emission suggesting maximum
temperatures less than 80000K if collisionally dominated.

\item The decay of stellar surface emission with decreasing
temperature for both C III and O VI  suggests that 
similar magnetic processes are responsible
for these emissions.

\item An outward acceleration of 80000K material, clearly  indicated
by the C III emission profiles occurs in all these
stars from F0 II through K5 III (\gal\ Tau) clearly demonstrating
the presence of a warm wind.

\item The O VI \gla 1032 emission gives some evidence also of wind
opacity in most stars suggesting that warmer winds
of 300000K may be present.

\item Semi-empirically modeling of atmospheres and 
winds and of the emergent chromospheric and transition region 
line profiles is  needed for luminous cool stars.

\end{enumerate}
\acknowledgements
This work is based on data obtained for the Guaranteed
Time Team by the NASA-CNES-CSA \fuse\ mission operated 
by the Johns Hopkins University.
Financial support to U.S. participants has been provided by
NASA Contract NAS5-32985.

\clearpage
\clearpage
\clearpage

\clearpage

\clearpage

\clearpage

\clearpage

\clearpage

\clearpage

\clearpage

\clearpage

\clearpage

\clearpage

\clearpage

\clearpage

\clearpage

\clearpage

\clearpage

\clearpage
\clearpage
\clearpage

\clearpage

\clearpage

\begin{deluxetable}{lccl}
\tablecaption{Major Atomic Transitions Considered Here\label{tbl.lines}}
\tablewidth{0pt}
\tablecolumns{4}

\tablehead{
\colhead{Ion}&\colhead{Transition\tablenotemark{a}}&\colhead{Wavelength (\AA)}
&\colhead{T$_{formation}$ (K)\tablenotemark{b}}

}
\startdata
Fe II& $many$\tablenotemark{c}   &1106--1143& 1.8 $\times$10$^4$ \\
C III& 2s$^2$2p $^1$S$_0$ -- 2s2p$^2$\ $^1$P$_1$&\phantom{1}977.02 &8.3 $\times$ 10$^4$ \\
C III& 2s2p $^3$P$_{0,1,2}$ -- 2p$^2$
$^3$P$_{0,1,2}$&1176.\tablenotemark{d}&8.3 $\times$ 10$^4$\\
O VI& 2s $^2$S$_{1/2}$ -- 2p $^2$P$_{3/2}$&1031.92&2.9 $\times$10$^5$  \\
O VI& 2s $^2$S$_{1/2}$ -- 2p $^2$P$_{1/2}$&1037.61& 2.9 $\times$10$^5$\\
Fe XVIII& 2p$^5$ $^2$P$_{3/2}$ -- 2p$^5$ $^2$P$_{1/2}$&\phantom{1}974.86&6.6
$\times$10$^6$\\
Fe XIX& 2p$^4$ $^3$P$_2$ -- 2p$^4$ $^3$P$_1$&1118.07&7.9
$\times$10$^6$\\

\enddata
\tablenotetext{a}{Atomic configuration expressed as lower level ($i$)
-- upper level ($j$) where emission corresponds to the
transition $j \rightarrow i$. }
\tablenotetext{b}{Temperatures correspond to the temperature
of peak emission rate in a collisionally dominated
plasma and were calculated with v.4 of the CHIANTI
database (Young \etal\ 2003) using the ionization balance calculations
of Mazzotta \etal\ (1998) and an electron density of 10$^{10}$
cm$^{-3}$.
}
\tablenotetext{c}{Fe II emission in these stars 
results from fluorescent and cascade processes involving many
configurations, and the  temperature of `formation'
does not strictly apply since the population of levels
is not linked to the local electron temperature.}
\tablenotetext{d}{There are six components to this multiplet.}
\end{deluxetable}


\begin{deluxetable}{lrcrrlrcl}
\tablecaption{Stellar Parameters\label{starparams}}
\tablewidth{0pt}
\tablecolumns{9}

\tablehead{
\colhead{Star}&\colhead{HD}&\colhead{Sp. Type}
&\colhead{Distance}&\colhead{V}&\colhead{V$-$R}&\colhead{\gal \tablenotemark{a}}
&\colhead{v$_{rad}$\tablenotemark{b}}&\colhead{References}\\
\colhead{}&\colhead{}&\colhead{}&\colhead{(pc)}&\colhead{}&\colhead{}
&\colhead{}&\colhead{(km s$^{-1}$)}
}
\startdata
\gb\ Cet &   4128& K0 III   &29.4& 2.02 &+0.72&6.74($+$15)&+13.4&1,2,3,5 \\
\gal\ Tau&  29139& K5 III   &20.0& 0.86&+1.23&3.09($+$14)&+54.0&1,2,3,4 \\
\gal\ Ori&  39801& M2 Iab   &131\phm{.0}& 0.42&+1.64 &5.31($+$13)&+21.0&1,2,3 \\
\gal\ Car&  45348& F0 II    &95.9& $-$0.75&+0.24&3.50($+$15)&+20.5&1,2,3 \\
\gb\ Gem &  62509& K0 IIIp  &10.3&1.15&+0.75&2.69($+$15)&+2.81&1,2,3,4 \\
31 Com   & 111812& G0 IIIp  &94.2&4.94&+0.55&1.94($+$17)&$-$1.25&1,2,3,4 \\
\gb\ Dra & 159181& G2Ib--IIa&110\phm{.0}&2.79&+0.68&1.60($+$16)&$-$21.6&1,2,3,4 \\
\gal\ Aqr& 209750& G2Ib     &$\sim$230\phm{.0}&2.95&+0.66&2.01($+$16)&+6.8&1,3,5\\
\enddata
\tablenotetext{a}{The factor, \gal, relates the flux observed at
Earth to the stellar surface flux $F_\star=\alpha\times
F_\oplus$ and $\alpha = (d/R_\star)^2=1.7018\times 10^{17}/\phi^2$
with $\phi$(mas) evaluated  from the Barnes-Evans relationships
Barnes \etal\ (1978). Numbers in parentheses denote the multiplying
power of 10 in the value of $\alpha$.}
\tablenotetext{b}{v$_{rad}$ denotes the heliocentric radial velocity
of the star (km s$^{-1}$).}
\tablecomments{Reference: (1) Perryman \etal\ 1997;
(2) GCRV 1953;
(3) Johnson \etal\ 1966; (4) deMedeiros \& Mayor 1999; (5) Beavers \&
Eitter 1986.}

\end{deluxetable}


\clearpage
\begin{deluxetable}{lrcllcc}
\tablecaption{Targets and FUSE Observation Log\label{exposures}}
\tablewidth{0pt}

\tablehead{
\colhead{Star}&\colhead{HD}&\colhead{Sp. Type}&\colhead{Dataset}&\colhead{Obs. Date}&
\colhead{Exposure ($ks$)}&\colhead{Apert.}
}
\startdata
\gb\ Cet& 4128&K0 III&P1180501&2000 Dec 10&13.1&LWRS \\
\gal\ Ori&39801&M2 Iab&P1180901&2000 Nov 3&10.3&LWRS\\
\gal\ Tau&29139&K5 III&P1040901&2001 Jan 14&12.2\tablenotemark{a}&MDRS\\
&\nodata&\nodata&P2180601&2003 Sep 14&6.3&LWRS\\
&\nodata&\nodata&P2180602&2003 Sep 15&12.2&LWRS\\
&\nodata&\nodata&P2180603&2003 Sep 15&10.4&LWRS\\
\gal\ Car&45348&F0 II&P1180101&2000 Dec 11&5.6&LWRS\\
&\nodata&\nodata&P2180101&2001 Oct 25&5.9&LWRS\\
&\nodata&\nodata&P2180102&2001 Oct 26&10.7&LWRS\\
\gb\ Gem&62509&K0 IIIp&P1180601&2000 Nov 11&21.8&LWRS\\
31 Com&111812&G0 IIIp&P1180401&2001 Apr 20&12.2&LWRS\\
\gb\ Dra&159181&G2Ib-IIa&P1180301&2000 May 9&5.6&LWRS\\
&\nodata&\nodata&P2180301&2001 Jun 30&16.4\tablenotemark{b}&LWRS\\
\gal\ Aqr&209750&G2 Ib&P2180201&2001 Jun 16&34.3&LWRS\\
&\nodata&\nodata&P2180202&2001 Oct 7&10.5&LWRS\\
\enddata
\tablenotetext{a} {The SiC 1B channels were on the target for  
a total of 1790 s and the SiC 2A exposure totaled 3182 s.}
\tablenotetext{b}{SiC 1B channel: Exposure 3 (1635 s) was not
on target.}
\end{deluxetable}


\clearpage

\begin{deluxetable}{lrlclc}
\tablecaption{Interstellar Medium Lines: Heliocentric Velocities\label{ism}}
\tablewidth{0pt}
\tablecolumns{5}

\tablehead{
\colhead{Star}&\colhead{HD}&\colhead{Species}&\colhead{$\lambda$}&
\colhead{$V_{ism}$}&\colhead{Note} \\
\colhead{}&\colhead{}&\colhead{}&\colhead{\it (\AA)}&\colhead{\it (km s$^{-1}$)}&
\colhead{}
}
\startdata
$\beta$ Ceti&4128& C II &1334& +5.8&1 \\
&\nodata& Si III& 1206&+4.5&1 \\
&\nodata&D I&1215&+4.7&1 \\
&\nodata&O I&1302&+6.6&1 \\
&&$Average\ Adopted$&\nodata&+5.4\\
$\alpha$ Tau&29139& Mg II&2795&--30.&2 \\
&\nodata&O I& 1302&--30.&2 \\
$\alpha$ Ori&39801&$not\ available$  \\
$\alpha$ Car&45348&H Ly-$\alpha$&1215.&+25.&3 \\
$\beta$ Gem&62509& Mg II ($Averaged$)&2800&+26.1&4 \\
31 Com&111812&Mg II&2800&--3.4&5 \\
&\nodata&Fe II&2599&--2.4&5\\
&\nodata&D I&1216&--2.7&5\\
&\nodata&C II&1335&--3.8&6\\
&\nodata&O I&1302&--2.6&6\\
&&$Average\ Adopted$&\nodata&--3.2\\
$\beta$ Dra&159181&Mg II &2800&--20.&7\\  
$\alpha$ Aqr&209750& Mg II&2800&--18&8 \\
\enddata
\tablecomments{\small (1) Measured from STIS spectrum (2)
Robinson  et al. 1998.  
(3) Marilli et al. 1997.
(4) Dring  et al. (1997) find 2 interstellar
components at +22.0$\pm$1.8~\kms\ and +33.2$\pm$1.8~\kms, of which
the 22 \kms\ cloud has a larger column density by a factor of 1.7; 
an average value, weighted by the column density is used, viz.:
$(N_1/N_{tot}) \times V_1 + (N_2/N_{tot}) \times V_2 = V_{avg}$.
(5) Values taken from Dring et al.; in agreement with independent
determination by Piskunov et al. (1997). (6) Redfield \&
Linsky (2004). (7) Measured
from GHRS spectra calibrated with Pt Lamp.
(8) IUE spectra from Drake et al. (1984). 
}
\end{deluxetable}


\clearpage


\begin{deluxetable}{llrlrlr}
\tablecaption{Emission Line Fluxes\tablenotemark{a}\label{tblflux}}
\tablewidth{0pt}
\tablecolumns{7}

\tablehead{
\colhead{Star}&\multicolumn{2}{c}{C III: $\lambda$977\tablenotemark{b}}
&\multicolumn{2}{c}{C III: $\lambda$1176\tablenotemark{c}}
&\multicolumn{2}{c}{O VI: $\lambda$1032\tablenotemark{d}}\\
\cline{2-3} \cline{4-5} \cline{6-7}\\
\colhead{}&\colhead{Flux}&\colhead{Counts}&\colhead{Flux}
&\colhead{Counts}&\colhead{Flux}&\colhead{Counts}
}
\startdata
\gb\ Cet&6.55($-$13)&4503&3.12($-$13)&4686&5.86($-$13)&10044\\
\gal\ Tau&7.00($-$14)&284&4.91($-$14)&518&3.34($-$14)&372\\
\gal\ Car&3.40($-$13)&4220&1.30($-$13)&6288&2.16($-$13)&6949\\
\gb\ Gem&2.90($-$13)&3357&6.95($-$14)&1780&1.07($-$13)&3177\\
31 Com&1.01($-$12)&7141&5.65($-$13)&8400&5.09($-$13)&8959\\
\gb\ Dra&1.52($-$12)&18203&4.54($-$13)&11694&8.03($-$13)&24133\\
\gal\ Aqr&2.77($-$13)&6226&7.75($-$14)&4233&1.04($-$13)&6188\\
\enddata
\tablenotetext{a}{Total flux ($erg\ cm^{-2}\ s^{-1}$) observed at
Earth in the emission line
obtained by integration over the line profile.  Errors in the
flux can be taken as $\pm$10\%
corresponding to the uncertainty in the FUSE absolute calibration. 
Numbers in parentheses denote the multiplying power of 10 
to be applied to the flux value.} 
\tablenotetext{b}{Measured from SiC2A channel.}
\tablenotetext{c}{Measured from LiF2A channel.}
\tablenotetext{d}{Measured from LiF1A channel.  A small background
continuum was subtracted from the flux values, although the
counts include the total of line plus continuum.}

\end{deluxetable}

\clearpage

\begin{deluxetable}{lcccccc}
\tablecaption{Gaussian Fits to Photon Spectrum of  C III \gla977 
\tablenotemark{a}\label{tbl.c3fits}}
\tablewidth{0pt}
\tablecolumns{7}

\tablehead{
\colhead{Star}&\multicolumn{3}{c}{Gaussian Fit (single)}
&\multicolumn{3}{c}{Gaussian Fit (long \gla\ side)}\\
\cline{2-4} \cline{5-7}\\
\colhead{}&\colhead{Vel$_{center}$}&\colhead{FWHM}
&\colhead{Flux\tablenotemark{b}$_{Gauss.}$}
&\colhead{Vel$_{center}$}&\colhead{FWHM}&\colhead{Flux\tablenotemark{b}$_{Gauss.}$}\\
\colhead{}&\colhead{($km\ s^{-1}$)}&\colhead{($km\ s^{-1}$)}&
\colhead{($erg\ cm^2\ s^{-1}$)}&
\colhead{($km\ s^{-1}$)}&\colhead{($km\ s^{-1}$)}&\colhead{($erg\ cm^2\ s^{-1}$)}
}
\startdata
\gb\ Cet&$+$15.6$\pm$3.1&131$\pm$2&6.02$\pm$0.6($-$13)&$-$1.85$\pm$3.0&143$\pm$2&7.98$\pm$0.8($-$13)\\
\gal\ Tau\tablenotemark{c}&$+$10.9$\pm$6.0&100$\pm$10&5.70$\pm$0.6($-$15)&\nodata&\nodata&\nodata\\
\gal\ Car&$+$11.6$\pm$0.9&180$\pm$3&3.60$\pm$0.4($-$13)&$-$2.60$\pm$2.5&184$\pm$3&4.72$\pm$0.5($-$13)\\
\gb\ Gem&$+$9.10$\pm$1.5&106$\pm$2&2.75$\pm$0.3($-$13)&$-$0.28$\pm$1.5&102$\pm$2&3.58$\pm$0.4($-$13)\\
31 Com&$+$20.3$\pm$1.5&250$\pm$4&1.04$\pm$0.1($-$12)&$-$4.30$\pm$2.0&281$\pm$4&1.26$\pm$0.2($-$12)\\
\gb\ Dra&$+$26.9$\pm$1.3&190$\pm$3&1.96$\pm$0.2($-$12)&$-$2.80$\pm$4.0&207$\pm$4&2.99$\pm$0.3($-$12)\\
\gal\ Aqr&$+$57.5$\pm$2.1&126$\pm$2&2.70$\pm$0.3($-$13)&$-$0.67$\pm$3.0&157$\pm$2&8.52$\pm$0.9($-$13)\\
\enddata
\tablenotetext{a}{Gaussian profiles fit to stellar emission profile
from SiC2A channel in photons (counts) using Poisson statistics (see text).}
\tablenotetext{b}{Fluxes as measured from the Gaussian fit, should
be used for relative contribution only. Numbers in parentheses denote
the multiplying power of 10 in the flux value.}
\tablenotetext{c}{Single Gaussian fit to rebinned data.}
\end{deluxetable}

\clearpage

\begin{deluxetable}{lcccccc}
\tablecaption{Parameters of 2 Gaussian Fits to O VI \gla1032\tablenotemark{a}\label{tblprofiles}}
\tablewidth{0pt}
\tablecolumns{7}

\tablehead{
\colhead{Star}&\multicolumn{3}{c}{2-Gaussian Fit (narrow)}
&\multicolumn{3}{c}{2-Gaussian Fit (wide)}\\
\cline{2-7}  \\
\colhead{}
&\colhead{Vel$_{center}$}
&\colhead{FWHM}
&\colhead{Flux\tablenotemark{b}$_{Gauss}$}
&\colhead{Vel$_{center}$}
&\colhead{FWHM}
&\colhead{Flux\tablenotemark{b}$_{Gauss}$}\\
\colhead{}&\colhead{($km\ s^{-1}$)}&\colhead{($km\ s^{-1}$)}
&\colhead{($erg\ cm^2\ s^{-1}$)}
&\colhead{($km\ s^{-1}$)}&\colhead{($km\ s^{-1}$)}
&\colhead{($erg\ cm^2\ s^{-1}$)}
}
\startdata
\gb\ Cet&$+$20.0$\pm$1.8&75$\pm$2&2.87$\pm$0.3($-$13)&$+$32.1$\pm$2.0&191$\pm$10&2.58$\pm$0.3($-$13)\\
\gal\ Tau\tablenotemark{c}&$-$1.45$\pm$2.0&182$\pm$6&2.47$\pm$0.2($-$14)&\nodata&\nodata&\nodata\\
\gal\ Car&$+$17.6$\pm$2.0&87$\pm$3&1.02$\pm$0.1($-$13)&$+$18.4$\pm$3.1&253$\pm$15&1.09$\pm$0.1($-$13)\\
\gb\ Gem&$+$0.142$\pm$1.5&75$\pm$2&7.57$\pm$0.8($-$14)&$-$7.76$\pm$2.1&157$\pm$15&2.96$\pm$0.3($-$14)\\
31 Com&$+$6.77$\pm$1.5&168$\pm$8&2.75$\pm$0.3($-$13)&$-$9.89$\pm$2.0&344$\pm$20&2.29$\pm$0.2($-$13)\\
\gb\ Dra&$+$15.2$\pm$2.0&137$\pm$5&4.97$\pm$0.5($-$13)&$+$16.2$\pm$1.9&239$\pm$15&2.95$\pm$0.3($-$13)\\
\gal\ Aqr&$+$2.48$\pm$1.5&99$\pm$4&2.79$\pm$0.3($-$14)&$-$5.46$\pm$2.0&210$\pm$10&6.91$\pm$0.7($-$14)\\
\enddata
\tablenotetext{a}{Gaussian profiles fit to stellar emission profile
from LiF1A in photons (counts) using Poisson statistics (see text).}
\tablenotetext{b}{Fluxes as measured from the Gaussian fit, should
be used for relative contribution only.  Numbers in parentheses
indicate the multiplying power of 10 in the flux value.}
\tablenotetext{c}{The O VI profile does not have sufficiently
good statistics to attempt a 2 Gaussian profile fit.}

\end{deluxetable}

\clearpage

\begin{deluxetable}{llcc}
\tablecaption{Parameters of Single Gaussian Fit to  O VI \gla1032\tablenotemark{a}\label{tblo6.red}}
\tablewidth{0pt}
\tablecolumns{4}

\tablehead{
\colhead{Star}&
\multicolumn{3}{c}{Gaussian Fit (long $\lambda$ side)}\\
\cline{2-4}  \\
\colhead{}&\colhead{Vel$_{center}$}&\colhead{FWHM}
&\colhead{Flux\tablenotemark{b}$_{Gauss}$}\\
\colhead{}&\colhead{($km\ s^{-1}$)}&\colhead{($km\ s^{-1}$)}
&\colhead{($erg\ cm^2\ s^{-1}$)}
}
\startdata
\gb\ Cet&$-$5.85$\pm$10&156$\pm$5&7.82$\pm$0.8($-$13)\\
\gal\ Car&$-$1.99$\pm$15&167$\pm$10&3.53$\pm$0.4($-$13)\\
\gb\ Gem&$-$7.55$\pm$6 &102$\pm$5 &1.27$\pm$0.1($-$13) \\
31 Com&$-$7.02$\pm$15 & 243$\pm$15 & 5.49$\pm$0.5($-$13)\\
\gb\ Dra&$+$1.38$\pm$10 & 182$\pm$8 &9.12$\pm$0.9($-$13) \\
\gal\ Aqr&$-$8.10$\pm$8 &189$\pm$9 &9.68$\pm$1.0($-$14) \\
\enddata
\tablenotetext{a}{Gaussian profile fit to long wavelength side
of  stellar emission profile
from LiF1A in photons (counts) using Poisson statistics (see text).}
\tablenotetext{b}{Fluxes as measured from the Gaussian fit, should
be used for relative contribution only.  Numbers in parentheses
indicate the multiplying power of 10 in the flux value.}

\end{deluxetable}



\begin{thebibliography}{}
\bibitem[]{}Achour, H., Brekke, P., Kjeldseth-Moe, O., \& Maltby,
P. 1995, ApJ, 453, 945
\bibitem[]{}Ake, T., B., Dupree, A. K., Young, P. R., Linsky, J. L.,
Malina, R. F., Griffiths, N. W., Siegmund, O. H. W., \&
Woodgate, B. E. 2000, ApJ, 538, L87
\bibitem[]{}Ayres, T. R., Brown, A., Harper, G. M., Osten, R. A.,
Linsky, J. L., Wood, B. E., \& Redfield, S. 2003, ApJ, 583, 963
\bibitem[]{}Ayres, T. R., Brown, A., Osten, R. A., Huenemoerder, D. P.,
Drake, J. J., Brickhouse, N. S., \& Linsky, J. L. 2001a, \apj,
549, 554
\bibitem[]{}Ayres, T. R., Osten, R. A., \& Brown, A. 1999, \apj, 526, 445
\bibitem[]{}Ayres, T. R., Osten, R. A., \& Brown, A. 2001b, \apj, 562,
L83
\bibitem[]{}Ayres, T. R., Simon, T., Stern, R. A., Drake, S. A., 
Wood, B. E. \& Brown, A. 1998, \apj, 496, 428
\bibitem[]{}Ayres, T. R. \etal\ 1995, \apjs, 96, 223
\bibitem[]{}Baliunas, S. L., Guinan, E. F., \& Dupree, A. K. 1984,
\apj, 282, 733
\bibitem[]{}Barnes, T. G., Evans, D. S., \& Moffett, T. J. 1978, MNRAS, 183, 285
\bibitem[]{}Beavers, W. I., \& Eitter, J. J. 1986, ApJS, 62, 147
\bibitem[]{}Bookbinder, J., Walter, F., \& Brown A. 1992, ASP Conf. Ser. 26, Seventh
Cambridge Workshop on Cool Stars, Stellar Systems, and the Sun, 
ed. M. Giampapa \& J. Bookbinder, (San Francisco: ASP), 27
\bibitem[]{}Brown, A., Deeney, B. D., Ayres, T. R., Veale, A., \&
Bennett, P. D. 1996, ApJS, 107, 263
\bibitem[]{}Buchholz, B., Ulmschneider, P., \& Cuntz, M. 1998,
\apj, 494, 700 
\bibitem[]{}Carpenter, K. C., Robinson, R. D., Harper, G. M.,
Bennett, R. D., Brown, A., \& Mullan, D. J. 1999, \apj, 521, 382
\bibitem[]{}Cash, W. 1979, \apj, 228, 939
\bibitem[]{}Curdt, W., Brekke, P., Feldman, U., Wilhelm, K.,
Dwivedi, B. N., Sch\"uhle, U., \& Lemaire, P. 2001, A\&A, 375, 591
\bibitem[]{}Del Zanna, G., Landini, M., \& Mason, H. 2002, A\&A, 385, 968
\bibitem[]{}deMedeiros, J. R., \& Mayor, M. 1999, A\&AS, 139, 433
\bibitem[]{}Dere, K. P., Landi, E., Young, P. R., \& DelZanna, G.
2001, \apjs, 134, 331
\bibitem[]{}Doschek, G. A., Bohlin, J. D., \& Feldman, U. 1976,
ApJ, 205, L177
\bibitem[]{}Doyle, J. G., \& McWhirter, R. W. P. 1980, \mnras\ 193, 947
\bibitem[]{}Drake, S. A., Brown, A., \& Linsky, J. L. 1984, ApJ, 284, 774
\bibitem[]{}Dring, A. R., Linsky, J., Murthy, J., Henry, R. C., Moos,
W., Vidal-Madjar, A., Audouze, J., \& Landsman, W. 1997, ApJ, 488, 760
\bibitem[]{}Dupree, A. K., \& Brickhouse, N. S. 1995, in IAU
Symp. 176: Poster Proceedings, ed. K. G. Strassmeier, (Vienna: Institut f\"ur 
Astronomie Universit\"at Vienna), 184
\bibitem[]{}Dupree, A. K., \& Brickhouse, N. S. 1998, \apj, 500, L33
\bibitem[]{}Dupree, A. K., Brickhouse, N. S., Doschek, G. A., Green,
J. C., \& Raymond, J. C. 1993, ApJ, 418, L41
\bibitem[]{}Dupree, A. K., Foukal, P. V., \& Jordan, C. 
1976, \apj, 209, 621
\bibitem[]{}Dupree, A. K., \& Reimers, D. 1989, in Exploring
the Universe with the IUE Satellite, ed. Y. Kondo, (Boston: Kluwer),
321
\bibitem[]{}Dupree, A. K., Whitney, B. A., \& Avrett, E. H. 1992, 
in ASP Conf. Ser. 26, Seventh Cambridge Workshop on Cool Stars, Stellar Systems, and
the Sun, ed. M. Giampapa \& J. Bookbinder, (San Francisco: ASP), 525
\bibitem[]{}Dupree, A. K., Whitney, B. A., \& Pasquini, L. 1999,
\apj, 520, 751
\bibitem[]{}Fekel, F. 1997, PASP, 109, 514
\bibitem[]{}Fulbright, J. P., \& Johnson, J. A. 2003, ApJ, 595, 1154
\bibitem[]{}FUSE Observers Guide, V4.0 2002, ed. B. G. Andersson et
al. ({\it http://fuse.pha.jhu/edu/support/guide/guide\_V4.0.html\#INRES})
\bibitem[]{}General Catalogue of Stellar Radial Velocities (GCRV), 1953,
(Carnegie Institution: Washington), Publication 601.
\bibitem[]{}G\'omez de Castro, A. I. 2002, MNRAS, 332, 409
\bibitem[]{}Harper, G. M. 2001, ASP Conf. Ser. 223, Eleventh Cambridge Workshop on
Cool Stars, Stellar Systems, and the Sun, ed. R. J. Garc\'ia L\'opez,
R. Rebolo, \& M. R. Zapatero Osorio, (San Francisco: ASP), 368
\bibitem[]{}Harper, G. M., Wilkinson, E., Brown, A., Jordan, C.,
\& Linsky, J. L. 2001, \apj, 551, 486
\bibitem[]{}Hartman, H., \& Johansson, S. 2000, \apj, A\&A, 359, 627
\bibitem[]{}Hartmann, L., Dupree, A. K., \& Raymond, J. C. 1980, \apj,
236, L143
\bibitem[]{}Hartmann, L., Dupree, A. K., \& Raymond, J. C. 1981, \apj,
246, 193
\bibitem[]{}Hummer, D., G., \& Rybicki, G. B. 1968, \apj, 153, L107
\bibitem[]{}H\"unsch, M., \& Schr\"oder, K.-P. 1996, A\&A, 309, L51
\bibitem[]{}H\"unsch, M., Schmitt, J.H.M.M., Schr\"oder, K.-P., \&
Reimers, D. 1996, A\&A, 310, 801
\bibitem[]{}H\"unsch, M., Schmitt, J.H.M.M., \& Voges, W. 1998,
A\&AS, 127, 251 
\bibitem[]{}Johnson, H. L., Iriarte, B., Mitchell, R. I.,
\& Wisniewskj, W. J. 1966, Comm. Lunar Plan. Lab., 4, 99
\bibitem[]{}Jordan, C., \& Linsky, J. L. 1989, in Exploring the
Universe with the IUE Satellite, ed. Y. Kondo, (Boston: Kluwer), 259
\bibitem[]{}Jordan, C., McMurry, A. D., Sim, S. A., \& Arulvel,
M. 2001, MNRAS, 322, L5
\bibitem[]{}Kaufman, V., \& Martin, W. C. 1989, J. Opt. Soc. Am., B6,
1769
\bibitem[]{}Lemaire, P., Emerich, C., Vial, J.-C., Curdt, W., 
Schuhle, U., \& Wilhelm, K. 2002, in Proc. SOHO-11: From 
Solar Minimum to Maximum, ESA SP-508,  219
\bibitem[]{}Lehner, N., Jenkins, E. B., Gry, C., Moos, H. W., Chayer,
P., \& Lacour, S. 2003, ApJ, 595, 858
\bibitem[]{}Linsky, J. L., \& Wood, B. E. 1994, ApJ, 430, 342
\bibitem[]{}Lobel, A. \& Dupree, A. K. 2001, \apj, 558, 815
\bibitem[]{}Luck, R. E., \& Lambert, D. L. 1985, ApJ, 298, 782
\bibitem[]{}Marilli, E., Catalano, S., Freire Ferrero, R.,
Gouttebroze, P., Bruhweiler, F., \& Talavara, A. 1997, A\&A, 317, 521
\bibitem[]{}Mazzotta, P., Mazzitelli, G., Colafrancesco, S., \& 
Vittorio, N. 1998, A\&AS, 133, 403
\bibitem[]{}McCandliss, S. R. 2003, PASP, 115, 651
\bibitem[]{}McMurry, A. D., \& Jordan, C. 2000, MNRAS, 313, 423
\bibitem[]{}Miralles, M.  P., Cranmer, S. R., \& Kohl, J. L. 2001,
\apj, 560, L193
\bibitem[]{}Moos, W. H. \etal\ 2000, \apj, 538, L1
\bibitem[]{}Nousek, J. A. \& Shue, D. R. 1989, \apj, 342, 1207
\bibitem[]{}Perryman, M. A. C. \etal\ 1997, A\&A, 323, L49
\bibitem[]{}Peter, H. 2001, A\&A, 374, 1108
\bibitem[]{}Peter, H. 2004, IAU Symp. 219, Stars as Suns: Activity
Evolution, and Planets, ed. A. K. Dupree \& A. O. Benz, (San Francisco: ASP) 575
\bibitem[]{}Peter, H., \& Brkovi\'c, A. 2003, A\&A, 403, 287
\bibitem[]{}Peter, H., \& Judge, P. G. 1999, \apj, 522, 1148
\bibitem[]{}Piskunov, N., Wood, B. E., Linsky, J. L., Dempsey, R. C.,
\& Ayres, T. R. 1997, ApJ, 474, 315
\bibitem[]{}Rao, L. M, Baliunas, S. L., Robinson, C. R., Frazer, J.,
Woodard, L., \& Donahue, R. A. 1993, ASP Conf. Ser. 45,  Luminous High Latitude Stars,
ed. D. D. Sasselov, (San Francisco: ASP),  300
\bibitem[]{}Redfield, S., \& Linsky, J. 2004, ApJ, 602, 776
\bibitem[]{}Redfield, S., Linsky, J. L., Ake, T. B., Ayres, T. R., 
Dupree, A. K., Robinson, R. D., Wood, B. E., \& Young, P. R. 
2002, ApJ, 581, 626
\bibitem[]{}Redfield, S., Ayres, T. R., Linsky, J. L., Ake, T. B.,
Dupree, A. K., Robinson, R. D., \& Young, P. R. 2003, ApJ, 585, 993
\bibitem[]{}Reimers, D. 1977, A\&A, 57, 395
\bibitem[]{}Reimers, D., H\"unsch, M., Schmitt, J. H. M. M., \&
Toussaint, F. 1996, A\&A, 310, 813
\bibitem[]{}Robinson, R. D., Carpenter, K. G., \& Brown, A. 1998, ApJ,
503, 396
\bibitem[]{}Robinson, R. D., Linsky, J. L., Woodgate, B. E., \& Timothy, J. G.
2001, ApJ, 554, 368
\bibitem[]{}Robinson, R. D. \etal\ 1992, ASP Conf. Ser. 26, 
Cool Stars, Stellar Systems, and
the Sun, ed. M. S. Giampapa \& J. A. Bookbinder, (San Francisco: ASP),  31
\bibitem[]{}Rutten, R. G. M., \& Pylyser, E. 1988, A\&A, 191, 227
\bibitem[]{}Sahnow, D. J. \etal\ 2000, \apj, 538, L7
\bibitem[]{}Sanz-Forcada, J., Brickhouse, N. S., \& Dupree, A. K.
2002, \apj, 570, 799
\bibitem[]{}Sanz-Forcada, J., Brickhouse, N. S., \& Dupree, A. K.
2003, \apjs,  145, 147
\bibitem[]{}Schaller, G., Schaerer, D., Meynet, G., \& Maeder, A.
1992, A\&AS, 96, 269
\bibitem[]{}Smith, R. K., Brickhouse, N. S., Liedahl, D. A., \&
Raymond, J. C.  2001, ApJ, 556, L91
\bibitem[]{}Strassmeier, K. G., Wash\"uttl, A., \& Rice, J. B. 1994,
IBVS, 3994, 1
\bibitem[]{}Teriaca, L., Banerjee, D., \& Doyle, J. G. 1999, A\&A,
349, 636
\bibitem[]{}Withbroe, G., \& Noyes, R. W. 1977, ARAA, 15, 363
\bibitem[]{}Wood, B. E., Linsky, J. L., \& Ayres, T. R. 1997, \apj,
478, 745 
\bibitem[]{}Young, P. R., Dupree, A. K., Wood, B. E., Redfield, S.,
Linsky, J. L., Ake, T. B., \& Moos, H. W. 2001, \apj, 555, L121
\bibitem[]{}Young, P. R., DelZanna, G., Landi, E., Dere, K. P.,
Mason, H. W., \& Landini, M. 2003, ApJS, 144, 135
\end{thebibliography}
\end{document}